\documentclass[11pt]{iopart}

\usepackage{amssymb}
\usepackage{rotating}
\usepackage{bbm}
\usepackage{color}
\usepackage{cite}

\usepackage{slashed}

\def\ra{\rightarrow}
\def\be{\begin{equation}}
\def\ee{\end{equation}}
\def\gs{\mathrel{
   \rlap{\raise 0.511ex \hbox{$>$}}{\lower 0.511ex \hbox{$\sim$}}}}
\def\ls{\mathrel{
   \rlap{\raise 0.511ex \hbox{$<$}}{\lower 0.511ex \hbox{$\sim$}}}}
\newcommand{\obb}{0\mbox{$\nu\beta\beta$}}

\newcommand{\onbb}{neutrinoless double beta decay}
\newcommand{\ba}{\begin{array}{c}}
\newcommand{\baz}{\begin{array}{cc}}
\newcommand{\bad}{\begin{array}{ccc}}
\newcommand{\bav}{\begin{array}{cccc}}
\newcommand{\bea}{\begin{equation} \begin{array}{c}}
\newcommand{\eea}{ \end{array} \end{equation}}
\newcommand{\ea}{\end{array}}
\newcommand{\D}{\displaystyle}
\newcommand{\dms}{\mbox{$\Delta m^2_{\odot}$}}
\newcommand{\dma}{\mbox{$\Delta m^2_{\rm A}$}}
\newcommand{\meff}{\mbox{$\langle m_{ee} \rangle$}}

\newcommand{\sss}{\sin^2 \theta_{12}}

\begin{document}

\title[Neutrinoless double beta decay and neutrino physics]
{Neutrinoless double beta decay and neutrino physics}

\author{Werner Rodejohann}

\address{Max-Planck-Institut f\"ur Kernphysik, 
Saupfercheckweg 1, 69117 Heidelberg, Germany}
\ead{werner.rodejohann@mpi-hd.mpg.de}

\begin{abstract}
The connection of neutrino physics with
neutrinoless double beta decay is reviewed. After presenting the
current status of the PMNS matrix and the theoretical background of
neutrino mass and lepton mixing, we will summarize the various
implications of neutrino physics for double beta decay. The influence of light sterile 
neutrinos and other exotic modifications of the three neutrino 
picture is also discussed.

\end{abstract}

%Uncomment for PACS numbers title message
%\pacs{00.00, 20.00, 42.10}
% Keywords required only for MST, PB, PMB, PM, JOA, JOB? 
%\vspace{2pc}
%\noindent{\it Keywords}: Article preparation, IOP journals
% Uncomment for Submitted to journal title message
\submitto{Focus issue on Double Beta Decay in \JPG}
% Comment out if separate title page not required
\maketitle

\tableofcontents

\section{\label{sec:intro}Introduction}

This contribution to the focus issue on Double Beta Decay deals with
the connection of neutrinoless double beta decay (\obb)
\cite{Furry:1939qr}, 
\be \label{eq:main}
(A,Z) \to (A,Z+2) + 2 e^- \, , 
\ee
to neutrino physics. Observation of neutrinoless double beta decay
would show that lepton number is violated, a finding that would be as
important as observation of baryon number violation, i.e.~proton 
decay. The huge importance of the process is the reason for extensive
experimental and theoretical activities, as summarized in various recent reviews \cite{Avignone:2007fu,Giuliani:2010zz,Rodejohann:2011mu,GomezCadenas:2011it,Elliott:2012sp,Bilenky:2012qi,Vergados:2012xy}. 
Indeed, the present time is
an exciting time for neutrinoless double beta 
decay. The strongest limit on the life-time stemmed  from
2001 \cite{KlapdorKleingrothaus:2000sn}, and was improved very
recently \cite{Auger:2012ar}. 
A sizable number of new experiments is already
running, under construction, or in the planing phase. Different 
isotopes and experimental techniques will be used, see table
\ref{tab:expts}. In case more than one experiment sees a signal, it
will be possible to increase the credibility of the claims, test the
nuclear matrix element calculations, and maybe even test the
underlying mechanism of \obb. Recent reviews on the experimental
aspects of \obb~can be found in
\cite{Avignone:2007fu,Giuliani:2010zz,GomezCadenas:2011it,Elliott:2012sp},
the focus issue contribution discussing this is by Zuber.

The fact that this contribution concentrates on neutrino physics 
needs to be stressed, because one always has to keep in mind that two
main possibilities for \obb~exist \cite{Rodejohann:2011mu}: 

\begin{enumerate}
\item {\it Standard Interpretation}: 

neutrinoless double beta decay is mediated by light and massive
Majorana neutrinos (the ones which oscillate) and all other mechanisms
potentially leading to \obb~give negligible or no contribution;

\item {\it Non-Standard Interpretations}:

neutrinoless double beta decay is mediated by some other lepton number
violating physics, and light massive
Majorana neutrinos (the ones which oscillate) potentially leading 
to \obb~give negligible or no contribution. 
\end{enumerate}
Here we will focus only on the standard interpretation of light neutrino
exchange. Massive neutrinos are firmly established, and the vast
majority of models and theories predicts neutrinos to be Majorana
particles \cite{Majorana:1937vz}. Therefore, an interpretation of neutrinoless double beta
decay experiments in terms of neutrino physics is presumably the best
motivated one. However, one always has to consider the possibility that 
other physics is at the origin of \obb. A recent review on the 
various beyond the Standard Model sources for \obb~can be found in
\cite{Rodejohann:2011mu}. Alternative interpretations will be dealt
with in the contribution by Deppisch, Hirsch and P\"as, ways to
distinguish the mechanisms will also be discussed by Fogli and Lisi. Note
that, as formulated by the Schechter-Valle theorem \cite{Schechter:1981bd}, any mechanism
will lead to a neutrino Majorana mass term. However, this mass is 
generated by a 4-loop diagram, and is therefore negligible, see
\cite{Duerr:2011zd} for an explicit calculation. \\

\begin{table}[pt]
{\scriptsize
\caption{\label{tab:expts}Categorization of running and 
planned experiments \cite{Giuliani:2010zz} and isotopes under
consideration. Low energy resolution corresponds to about 10 \% at the
expected peak, high energy resolution is 1 \% or less. }
\begin{tabular}{c|c|ccc|c} \br

Name & Isotope & \multicolumn{3}{c|}{source $=$ detector; calorimetric with} 

& source $\neq$ detector \\ 

 &      & high energy res. & low energy res.  & event

topology & event topology\\ \hline

AMoRE & $^{100}$Mo & \checkmark & -- & -- & -- \\ 

CANDLES & $^{48}$Ca & -- & \checkmark & -- & -- \\ 

 COBRA& $^{116}$Cd (and $^{130}$Te)  & -- &  -- & \checkmark &  -- \\ 

CUORE & $^{130}$Te  &\checkmark  &  -- &  -- &  -- \\

DCBA  & $^{82}$Se or $^{150}$Nd  &  -- &  -- &  -- & \checkmark \\ 

EXO & $^{136}$Xe  &  -- &  -- & \checkmark &  -- \\ 

GERDA& $^{76}$Ge & \checkmark &  -- &  -- &  -- \\ 

KamLAND-Zen & $^{136}$Xe &  -- & \checkmark &  -- &  -- \\ 

LUCIFER & $^{82}$Se or $^{100}$Mo or $^{116}$Cd  & 

\checkmark &  -- &  -- & -- \\ 

 MAJORANA & $^{76}$Ge & \checkmark &  -- &  -- &  -- \\ 

MOON & $^{82}$Se or $^{100}$Mo or $^{150}$Nd 

 &  -- &  -- &  -- & \checkmark \\ 

NEXT & $^{136}$Xe &  -- &  -- & \checkmark &  -- \\ 

SNO+ & $^{150}$Nd  &  -- & \checkmark &  -- &  -- \\ 

SuperNEMO& $^{82}$Se or $^{150}$Nd  &  -- &  -- &  -- & \checkmark \\ 

XMASS & $^{136}$Xe  &  -- & \checkmark &  -- &  -- \\
\br  
\end{tabular} }
\end{table} 

As it turns out, if the
three active neutrinos of the Standard Model are massive Majorana particles,
 observation of \obb~can give information on the neutrino mass
ordering, the Majorana CP phases, and the neutrino mass scale. For the
latter observable, it should be noted that two other ways to measure
neutrino mass exist, namely in ``Kurie-plot'' experiments
\cite{Otten:2008zz} such as
KATRIN, or via cosmological observations
\cite{Abazajian:2011dt}. Different assumptions go into these 
complementary methods, and an interesting discussion arises when 
signals in two or even all three types of measurements are 
established. There is even more to \obb: it contains ``flavor
information'', in the sense that a number of the many proposed flavor symmetry
models for lepton mixing has testable predictions for neutrino mass observables, which
can be used to test the models, or rule them out.  

A popular modification
of the standard neutrino picture is to add one or two light sterile neutrinos,
in order to explain various anomalous signals, such as the
LSND/MiniBooNE results, the reactor anomaly, or certain 
cosmological/astrophysical observations. See \cite{Abazajian:2012ys}
for a detailed description of the current situation. The presence of light
sterile neutrinos will strongly impact the standard discussion of \obb, and is
a good example on how the physics interpretation of \obb~can get
completely changed in case the underlying assumptions are wrong. 

Extracting precise physics results from \obb~requires precise determination of the nuclear
matrix elements, which currently is not the case. Nevertheless, recent
years saw a significant improvement of the calculations, and an extensive
experimental program was launched \cite{Zuber:2005fu} in order to
support and test the calculations (for the standard mechanism) as much as possible. These 
experimental activities are discussed in contributions by Ejiri and
Frekers, and by Freeman, Grabmayr and Schiffer. Nuclear physics aspects
will be dealt with by Faessler, Simkovic and Rodin, by Suhonen and
Civitarese, by Menendez, by Engel and by Vogel. \\

In what follows we will summarize  general and current aspects of
neutrino phenomenology and theory in section \ref{sec:nu}. The main
part of this contribution is section \ref{sec:main}, where we
aim to discuss the various aspects of double beta decay and standard neutrino
physics. Light sterile neutrinos are added to the standard picture in
section \ref{sec:sterile} before we conclude in section \ref{sec:concl}.

\section{\label{sec:nu}Neutrino physics}

\subsection{Neutrino mass and lepton mixing: theoretical origin}

The theory behind neutrino mass and lepton mixing has been reviewed
for instance in \cite{Mohapatra:2006gs}. The only assumption that is
necessary for what follows, is that neutrinos are Majorana
particles. A pragmatic way to achieve this is to accept the presence
of the unique dimension 5 operator, inversely proportional to the
scale $\Lambda$, at which new physics sets in \cite{Weinberg:1979sa}  
\be \label{eq:Leff}
{\cal L}_{\rm eff} = \frac 12 \frac{h_{\alpha\beta}}{\Lambda}
\overline{L_{\alpha}^c}  \, \tilde \Phi \, \tilde{\Phi}^T \, L_{\beta} 
 \stackrel{\rm EWSB}{\longrightarrow} \frac 12 \, (m_\nu)_{\alpha \beta} \,
\overline{\nu_{\alpha}^c} \, \nu_{\beta} \, . 
\ee
Charge conjugated spinors are denoted by the superscript '$^c$', 
$L_\alpha = (\nu_\alpha, \alpha)^T$ are the lepton doublets of
flavor $\alpha = e, \mu, \tau$ and $\Phi$ is the Higgs doublet with 
vacuum expectation value $v = 174$ GeV. $m_\nu$ is the neutrino mass
matrix, given as the Yukawa coupling matrix $h_{\alpha \beta}$ times
$v^2$ divided by the high energy scale $\Lambda$. 
As can be seen, after electroweak symmetry breaking a Majorana mass matrix $m_\nu$ of order
$v^2 /\Lambda$ is generated. With the typical mass scale of $m_\nu
\simeq 0.05$ eV, it follows that $\Lambda \simeq 10^{15}$ GeV.  

There are several ways to generate the Weinberg operator ${\cal L}_{\rm eff}$ in
terms of fundamental particles. There are three tree-level
possibilities, called type I, II and III seesaw, the numbering
accidently corresponding to their popularity. For the type I 
seesaw
\cite{Minkowski:1977sc,Yanagida:1979as,Mohapatra:1979ia,GellMann:1980vs}
one introduces right-handed neutrinos, which are weak singlets. One
can integrate out their Majorana mass matrix $M_R$, resulting in 
\be\label{eq:mnuseesaw}
m_\nu = - m_D^T \, M_R^{-1} \, m_D \, ,
\ee
where $m_D$ is a Dirac mass matrix in the term $\overline{N_R} \, m_D
\, \nu_L$, expected to be of the order of the other Standard Model
masses.  
The Weinberg operator is here realized with $\Lambda \simeq M_R$. 
The type II (or triplet) seesaw \cite{Magg:1980ut,Lazarides:1980nt,Mohapatra:1980yp}
requires scalar Higgs triplets, and the type III seesaw
\cite{Foot:1988aq} fermion triplets. Combinations of the three terms
are of course possible, as well as more evolved seesaw variants, such
as double, linear, or inverse seesaw. There are also radiative
mechanisms which induce a Majorana neutrino mass matrix via loop
diagrams, involving new particles \cite{Zee:1980ai,Zee:1985id,Babu:1988ki}. Very often the heavy (seesaw) 
messengers can generate the baryon asymmetry by their 
decays in the early Universe, the so-called leptogenesis mechanism
\cite{Fukugita:1986hr,Davidson:2008bu}.  Hence, establishing the
Majorana nature of neutrinos and the presence of CP violation in the
lepton sector would strengthen our belief in this already very
appealing mechanisms. Note however that a link between \obb~and the
baryon asymmetry is not guaranteed. The same is true for the
connection between low energy CP violation and leptogenesis:
leptogenesis is very well possible if the low energy CP phases are all
zero \cite{Pascoli:2003uh,Davidson:2007va}.

Whatever the origin of the dimension 5 operator, neutrinos are
Majorana particles:  $\nu_i^c \equiv C \, \bar{\nu}_i^T =
\nu_i$. Diagonalizing the mass matrix $m_\nu$ via 
\be \label{eq:mnu}
m_\nu = U^\ast \, m_\nu^{\rm diag} \, U^\dagger  ~,\mbox{ where } 
m_\nu^{\rm diag} = {\rm diag}(m_1, m_2, m_3) \, , 
\ee
results for the charged current term in the appearance of 
the Pontecorvo-Maki-Nakagawa-Sakata (PMNS) 
matrix $U$: 
\be
{\cal L}_{CC} = -\frac{g}{\sqrt{2}} \, \overline{\ell}_\alpha \,
\gamma^\mu \, U_{\alpha i} \, \nu_i  \, W^-_\mu \, . 
\ee 
We have gone here without loss of generality in the basis in which the charged lepton mass matrix
is real and diagonal. The neutrino mass states
$\nu_{1,2,3}$ are superpositions of neutrino flavor states $\nu_{e, \mu, \tau}$: 
\be
\nu_\alpha = U_{\alpha i}^\ast \, \nu_i \, .
\ee
In the next subsection the observational status of the parameters in
$m_\nu$ will be discussed.

\subsection{\label{sec:mnu_obs}Neutrino mass and lepton mixing: observational status}

Since the mass matrix is complex and symmetric there are 9 physical
parameters in $m_\nu$, usually parameterized as three masses, three
angles and three phases. The PMNS mixing matrix $U$ is unitary and can
be written in its standard parametrization adopted by the PDG
\cite{Nakamura:2010zzi} as  
\be \label{eq:U}
U = \left( \bad 
c_{12}   c_{13} 
& s_{12}  c_{13} 
& s_{13}  e^{-i \delta}  \\ 
-s_{12}  c_{23} 
- c_{12}  s_{23} \, 
s_{13}   e^{i \delta} 
& c_{12}  c_{23} - 
s_{12}  s_{23}  s_{13} 
\, e^{i \delta} 
& s_{23}   c_{13}  \\ 
s_{12}    s_{23} - c_{12} 
 c_{23}  s_{13}  e^{i \delta} & 
- c_{12}  s_{23} 
- s_{12}  c_{23} \, 
s_{13}  e^{i \delta} 
& c_{23}   c_{13}  %\nonumber 
\ea   
\right) P \,,
\ee
where $s_{ij} = \sin \theta_{ij}$, $c_{ij} = \cos \theta_{ij}$ and
$\delta$ is the ``Dirac phase'' responsible for CP violation in
neutrino oscillation experiments. 
In eq.~(\ref{eq:U}) we
have included a diagonal phase matrix $P$, containing the two ``Majorana
phases'' $\alpha$ and $\beta$: 
\be\label{eq:P}
P = {\rm diag}(1,e^{i \alpha}, e^{i (\beta + \delta)}) \, . 
\ee
These phases are physical only if neutrinos are Majorana 
particles \cite{Bilenky:1980cx,Schechter:1980gr,Doi:1980yb}. We have
included here $\delta$ in $P$, as a consequence the $U_{ei}$ elements
are independent of $\delta$. The above parametrization is the product
of a 23-, 13- and 12-rotation, then multiplied with $P$, with a phase
in the 13-rotation. A ``symmetrical parametrization'' is also possible, in which each
individual rotation matrix contains a phase
\cite{Schechter:1980gr}. The resulting PMNS matrix is \cite{Rodejohann:2011vc}
\begin{equation}
\hspace{-2.6cm}U=\left( \begin{array}{c c c}
c_{12}c_{13}&s_{12}c_{13}e^{-i{\phi_{12}}}&s_{13}e^{-i{\phi_{13}}}\\
-s_{12}c_{23}e^{i{\phi_{12}}}-c_{12}s_{13}s_{23}e^{-i({\phi_{23}}-{\phi_{13}})}
&c_{12}c_{23}-s_{12}s_{13}s_{23}
e^{-i({\phi_{12}}+{\phi_{23}}-{\phi_{13}})}&c_{13}s_{23}e^{-i{\phi_{23}}}\\
s_{12}s_{23}e^{i({\phi_{12}}+{\phi_{23}})}-c_{12}s_{13}c_{23}e^{i{\phi_{13}}}
&-c_{12}s_{23}e^{i{\phi_{23}}}-
s_{12}s_{13}c_{23}e^{-i({\phi_{12}}-{\phi_{13}})}&c_{13}c_{23}\\
\end{array} \right) 
\label{writeout}
\end{equation} 
and its differences to the standard parametrization will be discussed
whenever appropriate.  

\begin{table}[t]
\caption{\label{tab:osc}Global fit results from 
neutrino oscillation experiments, taken from \protect \cite{Tortola:2012te}. The
values in brackets are for the inverted ordering. } 
\begin{indented}
\item[]\begin{tabular}{@{}cccc@{}}\br 
parameter    	&	best-fit$^{+1\sigma}_{-1\sigma}$ & $2\sigma$ &
$3\sigma$ \\\hline
$\Delta m^2_{21}\, \left[ 10^{-5}\, {\rm eV}^2 \right]$  &
$7.62^{+0.19}_{-0.19} $ & 7.27 -- 8.01 & 7.12 -- 8.20 \\ \hline 
$|\Delta m^2_{31}| \, \left[ 10^{-3}\, {\rm eV}^2 \right]$ & 
$2.53^{+0.08}_{-0.10} $ & 2.34 -- 2.69 & 2.26 -- 2.77 \\
& $\left(2.40^{+0.10}_{-0.07}\right) $ & (2.25 -- 2.59) & (2.15 --
2.68) \\  \hline 
$\sin^2 \theta_{12}$  & $0.320_{-0.017}^{+0.015}$ & 0.29 -- 0.35 & 0.27
-- 0.37 \\ \hline 
$\sin^2 \theta_{23}$ & $0.49_{-0.05}^{+0.08}$ & 0.41 -- 0.62 & 0.39 --
0.64 \\ 
& $\left(0.53^{+0.05}_{-0.07}\right) $ & (0.42 -- 0.62)  & (0.39 --
0.64)  \\ \hline 
$\sin^2 \theta_{13}$ & $0.026_{-0.004}^{+0.003}$ & 0.019 -- 0.033 & 0.015 --
0.036 \\ 
& $\left(0.027^{+0.003}_{-0.004}\right) $ & (0.020 -- 0.034)  & (0.016 -- 0.037)  \\ \hline 
$\delta/\pi$ & $0.83_{-0.64}^{+0.54}$ & 0 -- 2 & 0 -- 2
\\ 
& $-0.07$ & 0 -- 2 &  0 -- 2 
\\ 
 \br
\end{tabular}
\end{indented}
\end{table}

Neutrino oscillation experiments can probe in principle 6 of the nine
parameters in $m_\nu$, all angles, one phase\footnote{In the
symmetrical parametrization (\ref{writeout}) the phase combination $\phi_{13} -
\phi_{12} - \phi_{23}$ governs CP violation in oscillations, and one
immediately recognizes CP violation as a three-generation phenomenon, 
involving the phases of all three generations.} and two mass-squared
differences (including their sign). Fitting \cite{Tortola:2012te} the general formula 
\bea \nonumber
P(\nu_\alpha \to \nu_\beta) = \delta_{\alpha\beta} - 4 \sum\limits_{i>j}{\rm
  Re} \left\{  U_{\alpha i}^\ast \, U_{\beta j}^\ast \, U_{\beta i} \,
  U_{\alpha j} \right\} \sin^2 \frac{\Delta m^2_{ij} \, L}{4 \, E} \\
 + 2 \sum\limits_{i>j}{\rm
  Im} \left\{  U_{\alpha i}^\ast \, U_{\beta j}^\ast \, U_{\beta i} \,
  U_{\alpha j} \right\} \sin \frac{\Delta m^2_{ij} \, L}{2 \, E} \, , 
\eea
with $E$ the neutrino energy, $L$ the baseline, and including
matter effects whenever necessary, to the results of various neutrino oscillation
experiments gives the allowed ranges in table \ref{tab:osc}. Regarding
the mixing angle $\theta_{13}$, or the mixing matrix element
$|U_{e3}|$, there has recently been spectacular progress. Following
T2K \cite{Abe:2011sj}, all three reactor experiments have shown
evidence for non-zero and sizable $|U_{e3}|$, namely Double Chooz
\cite{Abe:2011fz}, Daya Bay \cite{An:2012eh} and RENO
\cite{Ahn:2012nd}: 
\be \label{eq:t13}
\begin{array}{ccc}
\mbox{Double Chooz:} & \sin^2 2 \theta_{13} = 0.086 \pm 0.051 & \neq 0\mbox{ at
} 1.9 \sigma\, ,  \\
\mbox{Daya Bay:} & \sin^2 2 \theta_{13} = 0.092 \pm 0.017 & \neq 0 \mbox{ at
} 5.2 \sigma\, ,  \\
\rm{RENO:} & \sin^2 2 \theta_{13} = 0.113 \pm 0.023 & \neq 0 \mbox{ at
} 4.9\sigma\, .
\end{array}
\ee
Combining the reactor data\footnote{At the Neutrino 2012 conference in
June 2012, Double Chooz have presented new data with $3.1\sigma$
evidence for non-zero $U_{e3}$, and also Daya Bay has increased its
significance to more than $7\sigma$. Also the T2K results were
updated.} with the other information rules out
vanishing $U_{e3}$ at more than $7\sigma$. 
As we will see, the precise value of $\theta_{13}$ is not particularly important for \obb, at least
for the current and next generation of experiments. Of more importance
for \obb~is the sign of the atmospheric mass-squared difference
$\Delta m^2_{31}$ (see section \ref{sec:NHIH}), and the neutrino mass
scale (section \ref{sec:mnu}), both are currently
unknown. An often overlooked issue for neutrinoless double beta decay
is the value of $\theta_{12}$, which is the best known mixing angle,
but still induces a sizable uncertainty for life-time predictions in
the inverted hierarchy \cite{Dueck:2011hu}, 
comparable to the nuclear physics uncertainty, as we will discuss in
section \ref{sec:NHIH}.

The neutrino mass ordering is called normal (inverted) if the atmospheric
mass-squared difference is larger (smaller) than zero: 
\be
\label{eq:masses}
\bad
\rm{normal:}  &  m_2 = \sqrt{m_1^2+\dms} ~;~~  m_3 =
\sqrt{m_1^{2}+\dma} \, ,\\
\rm{inverted: }~&  m_2 = \sqrt{m_3^2+\dms+\dma} ~;~~m_1 =
\sqrt{m_3^2 + \dma}  \, .
\ea
\ee
%\begin{figure}[t]
%\begin{center}
%\includegraphics[width=4cm,height=4cm]{SpettroDiretto2.ps}\hspace{2cm}
%\includegraphics[width=4cm,height=4cm]{SpettroInverso2.ps}
%\vspace*{8pt}
%\caption{\label{fig:NHIH}Normal and inverted mass ordering. The red
%  area denotes the electron content $|U_{ei}|^2$ in the mass state 
%  $\nu_i$, the yellow and blue areas denote the muon and
%  tau contents. Taken from \protect \cite{Strumia:2006db}.}
%\end{center}
%\end{figure}
With unknown mass ordering and total mass scale there are three
extreme cases, namely 
\be
\baz  
\mbox{ normal hierarchy~(NH): }  &  
   m_3 \simeq \sqrt{\dma} \gg m_{2} \simeq \sqrt{\dms} \gg m_1\,,\\[0.3cm]
\mbox{ inverted hierarchy~(IH): } &  
 m_2 \simeq m_1 \simeq \sqrt{\dma} \gg m_{3} \, ,\\[0.3cm]
\mbox{ quasi-degeneracy~(QD): } &  
  m_0^2 \equiv m_1^2 \simeq m_2^2 \simeq m_3^2  \gg \dma \,.
\label{eq:mass}
\ea
\ee 
The neutrino mass scale can be measured with three complementary
methods. The most model-independent one is by examining electron
spectra in $\beta$-decays, i.e.~{\bf Kurie-plot experiments} \cite{Otten:2008zz}. 
The observable neutrino mass parameter is 
\be \label{eq:mbeta}
m_\beta \equiv \sqrt{\sum |U_{ei}|^2 \, m_i^2 } \, . 
\ee
The current limit at 95\% C.L.~to this quantity from spectrometer approaches
is 2.3 eV and 2.1 eV, obtained from the Mainz \cite{Kraus:2004zw} and
Troitsk \cite{Lobashev:2003kt,Aseev:2011dq} collaborations, respectively.  
The KATRIN experiment \cite{Osipowicz:2001sq,Host:2007wh} has a discovery potential of
$m_\beta = 0.35$ eV with $5\sigma$ significance and a design sensitivity of 
$m_\beta = 0.2$ eV (90\% C.L.). It has been shown that neutrino mass
determination is very robust with respect to new physics
\cite{SejersenRiis:2011sj}. Such kind of experiments have reached
their ultimate size, and going lower in mass requires alternative
approaches. In principle, MARE (using crystal bolometers)
\cite{Monfardini:2005dk} and Project 8 (measuring cyclotron radiation
emitted by electrons) \cite{Monreal:2009za} can reach limits of 0.1
eV.

Neutrino mass determination via {\bf cosmological and astrophysical
observations} sets constraints on the sum of masses 
\be \label{eq:msum}
\Sigma = \sum m_i \, . 
\ee
Limits depend on the fitted data sets and the underlying cosmological 
model applied to the data, see
\cite{Hannestad:2010kz,Abazajian:2011dt} for recent
summaries. Therefore, it is difficult to set a robust and
model-independent constraint, upper limits on the sum of masses range
from 1.5 eV to less than 0.5 eV. It is fair to say that if indeed
neutrino masses are such that $\Sigma \gs 1$ eV, a rather unusual
cosmological model would be needed. 

The third approach to neutrino mass is {\bf neutrinoless double beta
decay}, which we will discuss in section \ref{sec:mnu}. Obviously, in
order to extract neutrino mass from \obb~one needs to assume that
neutrinos are Majorana particles, and are the leading contribution to
the process. The expression on which the life-time of \obb~depends is 
\be \label{eq:meff0}
\meff = \left| \sum U_{ei}^2 \, m_i \right| , 
\ee
and upper limits on $\meff$ can be translated in upper limits on
$m_\beta$ and $\Sigma$. 

Cosmology gives the strongest limits on neutrino mass, \obb~and direct
searches give comparable limits, see section \ref{sec:mnu}. However, we stress
again the conceptual differences of the different approaches to
neutrino mass, to be further discussed in section \ref{sec:mnu}.

\section{\label{sec:main}Neutrinoless double beta decay and neutrino physics}
Having set the stage, we can discuss the connection of neutrino
physics and neutrinoless double beta decay. 

\subsection{\label{sec:gen}General aspects} 
In general, the decay rate of \obb~can be written as  
\be \label{eq:fact}
\Gamma^{0\nu} = \sum_x G_x(Q,Z) \, |{\cal M}_x(A,Z) \, \eta_x|^2 \, ,
\ee
where the subscript $x$ denotes the underlying mechanism and $\eta_x$ are
functions of the particle physics parameters  
responsible for the decay. The nuclear matrix elements 
${\cal M}_x(A,Z)$ depend on the mechanism and the nucleus. Finally, 
$G_x(Q,Z)$ are phase space factors (depending on the $Q$-value with
$Q^5$ for most mechanisms including light neutrino exchange)  
which can have dependence on 
the particle physics.  Note that the possibility of destructive or
constructive interference of different mechanisms is present. However,
here we are only interested in the presence of one particle physics
mechanism, the exchange of light massive Majorana neutrinos, the ones
which are responsible for neutrino oscillations.

\begin{figure}[t]
\begin{center}
\includegraphics[width=6cm,height=4cm]{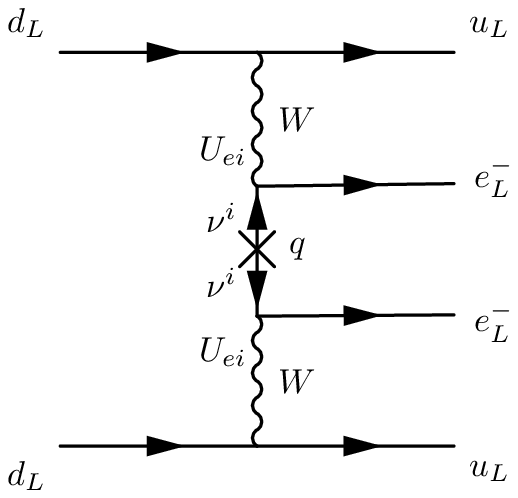}\quad
\includegraphics[width=5cm]{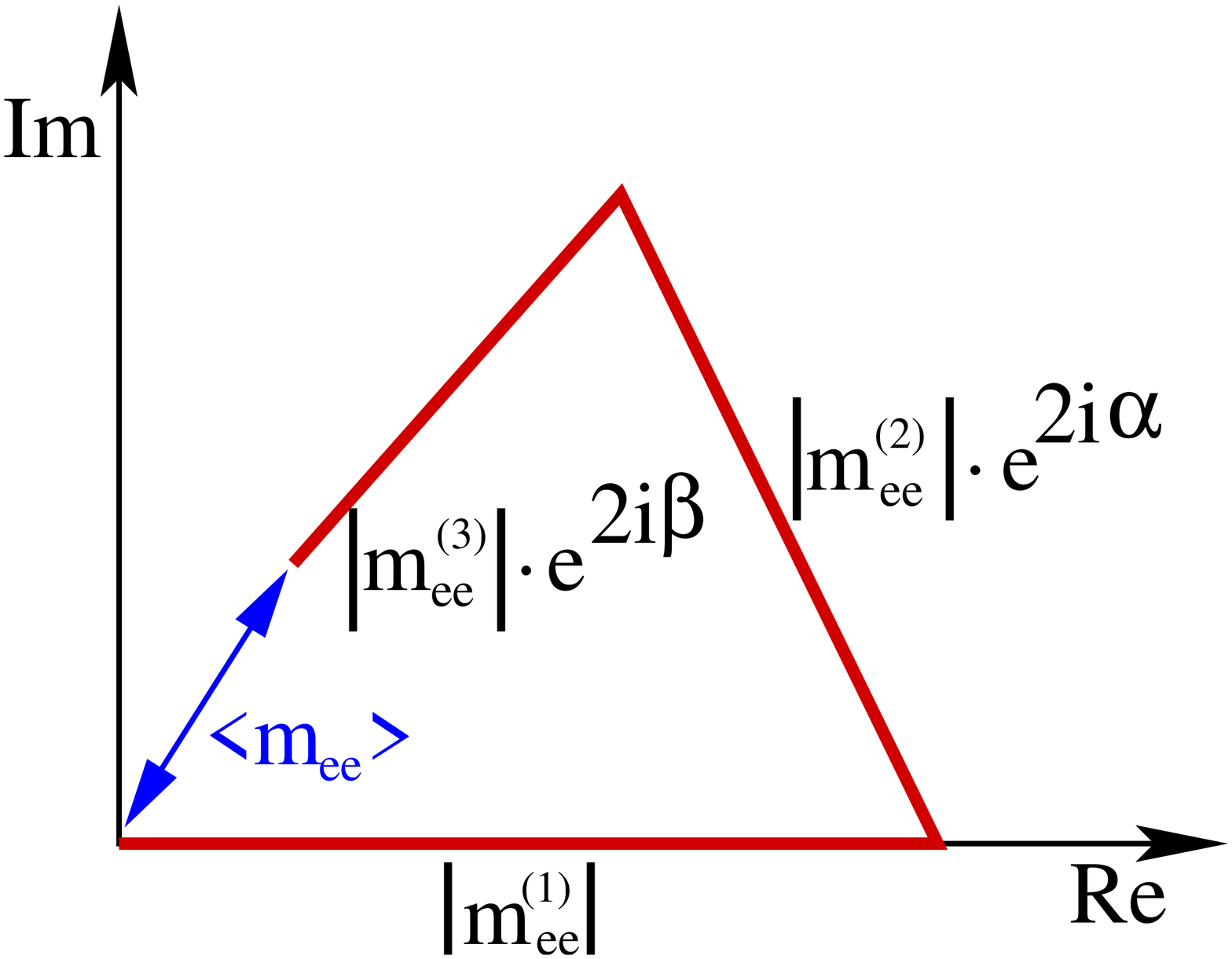}
\end{center}
\vspace*{8pt}
\caption{\label{fig:FD_mass_mech}Left: quark level ``lobster'' 
diagram for \onbb~in case of light Majorana neutrino exchange. Right: 
geometrical visualization of the effective mass. }
\end{figure}

The Feynman diagram for \obb~on the quark level in this
interpretation is shown in figure \ref{fig:FD_mass_mech}.  
The amplitude of the process is for the $V - A$ interactions of the
Standard Model proportional to  
\[ 
%bea \D  \label{eq:SA_calc}
\hspace{-2.5cm}\sum G_F^2 \, U_{ei}^2 \, \gamma_\mu \, 
\gamma_+ \frac{\slashed{q} + m_i}{q^2 - m_i^2} \, \gamma_\nu \, \gamma_-
= \sum G_F^2 \, U_{ei}^2 \frac{m_i}{q^2 - m_i^2} \gamma_\mu \,
\gamma_+ \, \gamma_\nu %\\ \D 
\simeq  \sum G_F^2 \, U_{ei}^2 \frac{m_i}{q^2} \gamma_\mu \, \gamma_+ \,
\gamma_\nu \, , 
\] %eea
where $\gamma_{\pm} = \frac12 (1 \pm \gamma_5)$, $m_i$ is the neutrino
mass, $q \simeq 100$ MeV is the typical neutrino momentum
(corresponding to the typical nuclear distance of about 1 fm), and $U_{ei}$ an
element of the first row of the PMNS matrix. 
The linear dependence on the neutrino mass is expected from the
requirement of a spin-flip, as the neutrino can be thought of as being emitted as a
right-handed state and absorbed as a left-handed state. 

The decay width of \obb~is therefore proportional to the square of the
so-called effective mass  
\be \label{eq:meff}
\meff = \left| \sum U_{ei}^2 \, m_i \right| 
= \left| |m_{ee}^{(1)}| + |m_{ee}^{(2)}| \, e^{2i\alpha} + 
|m_{ee}^{(3)}| \, e^{2i\beta} \right|  , 
\ee
which is visualized in figure \ref{fig:FD_mass_mech} (right) as the sum of three 
complex vectors $m_{ee}^{(1,2,3)}$. The effective mass is a coherent sum, which
implies the possibility of cancellations. Note that since we have
included $\delta$ in $P$ (see (\ref{eq:P})), the Dirac phase does not
appear in \meff, which is the way it should be. 
In the symmetrical parametrization (\ref{writeout}) the effective mass is given as 
\begin{equation} \label{eq:meff_expl}
\meff = \left| c_{12}^2 c_{13}^2  \,m_1 + s_{12}^2 c_{13}^2  \,m_2 \,e^{2 i \phi_{12}}+
s_{13}^2  \, m_3  \,e^{2 i \phi_{13}}
\right|  ,
\end{equation}
so only the two Majorana phases appear in
\meff~\cite{Rodejohann:2011vc}. 

Neutrinoless double beta decay is suppressed by the extremely small
ratio of neutrino mass $m_i \ls 0.5$ eV and momentum transfer $|q|
\simeq 10^8$ eV, and therefore only Avogadro's number provides a chance to 
create situations in which the Majorana nature of neutrinos is
observable. Compare this for instance with the decay $K^- \to \pi^+ \, e^- e^-$, which has a similar
momentum scale, and has an extremely tiny branching ratio of about 
\be
{\rm BR}(K^- \to \pi^+ \, e^- e^-) \sim 10^{-33} \,
\left(\frac{\meff}{\rm eV} \right)^2 \, , 
\ee
to be compared with the experimental upper limit of $6.4 \times 10^{-10}$ \cite{Appel:2000tc}.

Turning to experimental aspects, it is important to note that the life-time reach that
can be obtained in an experiment depends on the background level: 
\be \label{eq:Texp}
(T_{1/2}^{0\nu})^{-1} \propto \left\{ 
\baz \D 
a \, M \, \varepsilon \, t & \mbox{without background,} \\ \D 
a \, \varepsilon \, \sqrt{\frac{M \, t}{B \, \Delta E}} &
\mbox{with background.} 
\ea 
\right. 
\ee
Here $M$ is the mass, $t$ measurement time, $B$ background index with natural
units of counts/(keV kg yr) and $\Delta E$ the energy resolution at the
peak. Noting that the decay width depends quadratically on the
particle physics parameter, it is clear that within background
dominated experiments an improvement of the particle physics by a
factor 2 implies a highly non-trivial combined improvement of 16 on background index,
energy resolution,  mass and measurement time. 

There are similar (and
more difficult to observe) processes called neutrino-less double
beta$^+$ decay $(A,Z) \ra (A,Z-2) + 2 \, e^+ $
$(0\nu\beta^+\beta^+)$, beta$^+$-decay electron capture $e_b^- +
(A,Z) \ra (A,Z-2) + e^+$ 
$(0\nu\beta^+\rm EC)$, or double electron capture $2 \, e_b^- + (A,Z)
\ra (A,Z-2)^\ast$ $(0\nu\rm
ECEC)$ of bound state electrons $e^-_b$ (discussed in this focus issue
by Blaum, Eliseev and Novikov), which can also be searched for. 
Observation of one of those processes would also imply the
non-conservation of lepton number. The rates depend on
the particle physics parameters in the same way as \obb~does, we can
therefore focus on $\obb$. 

\begin{figure}[t]
\begin{center}
\includegraphics[width=7cm,angle=270]{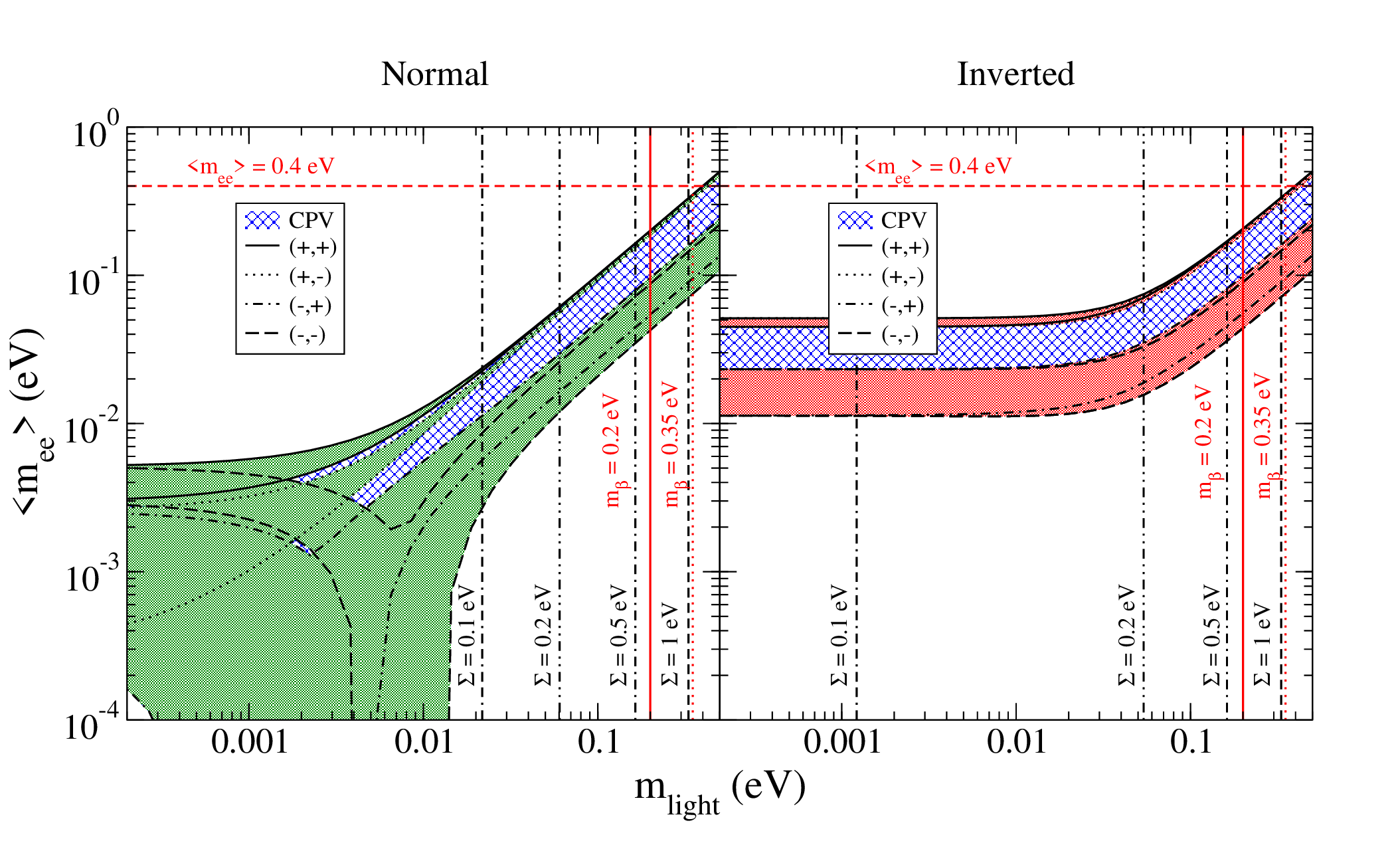}

\includegraphics[width=7cm,angle=270]{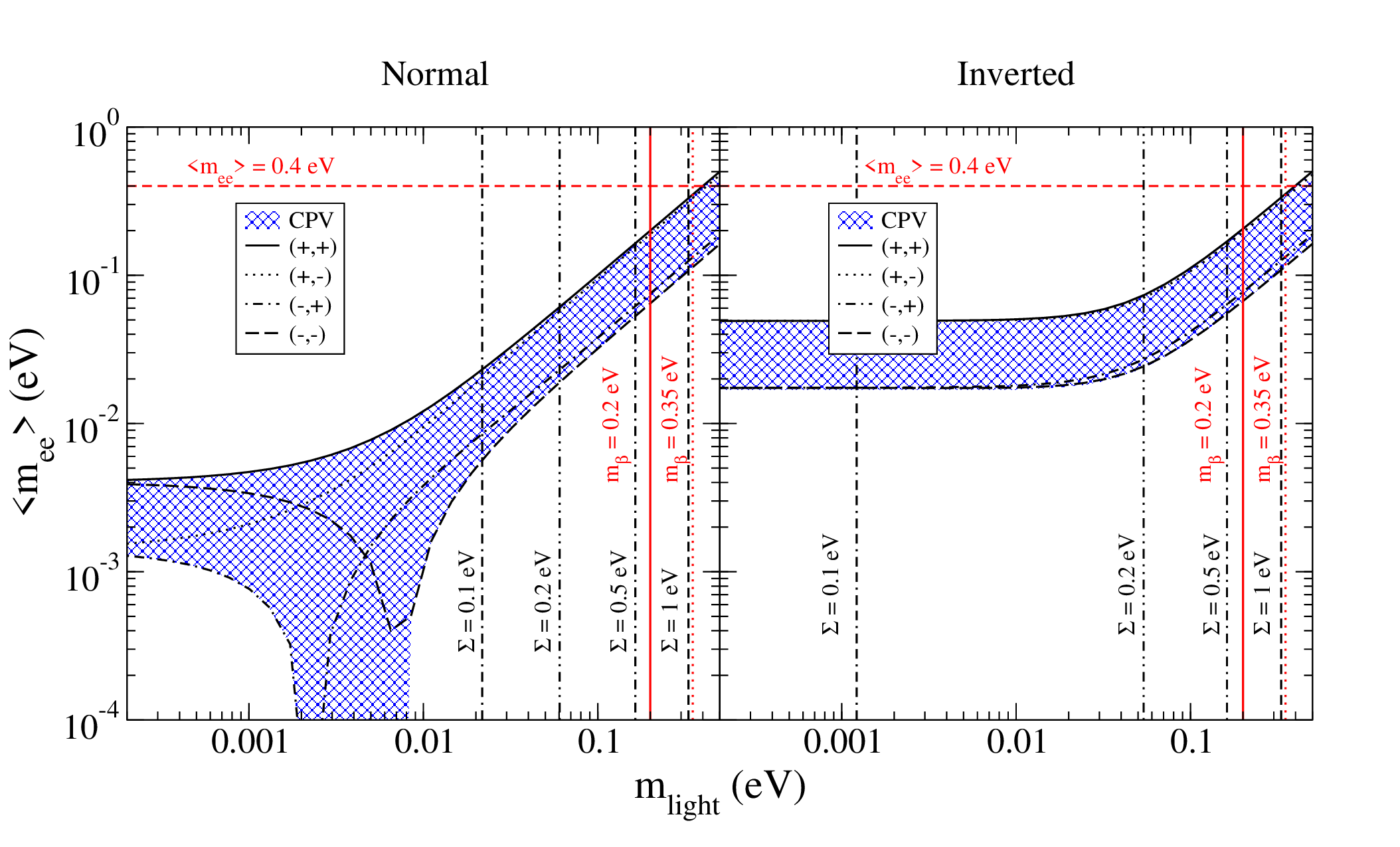}
\end{center}
\vspace*{8pt}
\caption{\label{fig:meff_mass}Effective mass against the smallest
neutrino mass for the $3\sigma$ ranges (top) and best-fit values
(bottom) of the oscillation parameters. The green and red shaded areas are the
general $3\sigma$ ranges, while the blue shaded areas can only be
realized if the CP phases take non-trivial values. $(\pm,\pm)$ denote different CP
conserving situations, corresponding to signs of $m_2$ and $m_3$,
relative to positive $m_1$. Prospective future values of $\Sigma$ and
$m_\beta$ are also given.}
\end{figure}

The effective mass depends on 7 out of
the 9 physical parameters of low energy neutrino physics (only
$\theta_{23}$ and the Dirac phase do not 
appear), hence contains an enormous amount of
information. It is the only realistic observable in which the two
Majorana phases appear. For the other five parameters there will be
complementary information from oscillation experiments or other
experiments probing neutrino mass. 
It is also noteworthy that \meff~is the $ee$ element of the
neutrino mass matrix $m_\nu$, see eq.~(\ref{eq:mnu}), which is a
fundamental object in the low energy Lagrangian. In terms of the
origin of neutrino mass, \meff~is $h_{ee} \, v^2/\Lambda$, see
eq.~(\ref{eq:Leff}). 

\begin{figure}[t]
\begin{center}
\includegraphics[width=8.cm,height=12cm,angle=270]{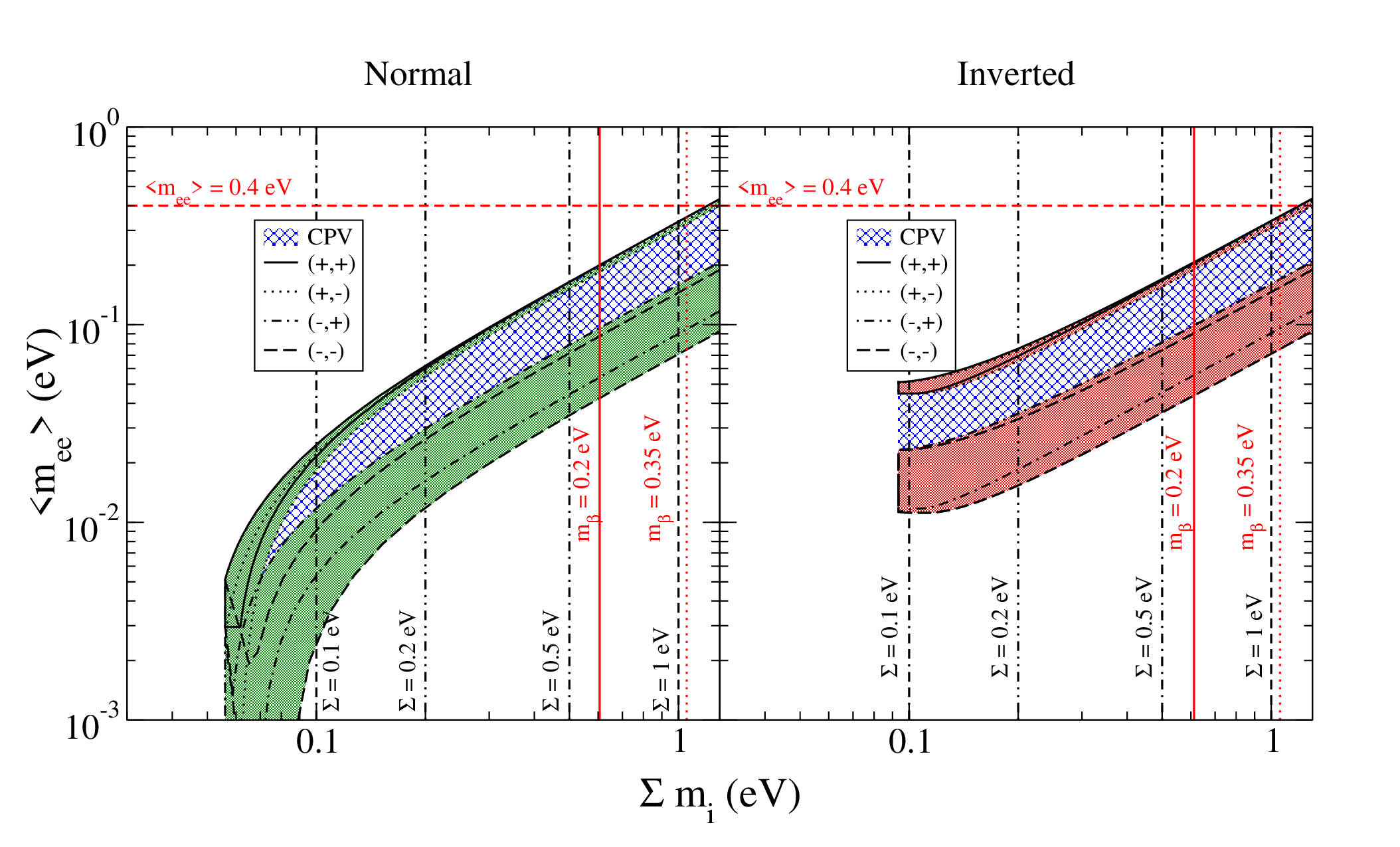}

\includegraphics[width=8.cm,height=12cm,angle=270]{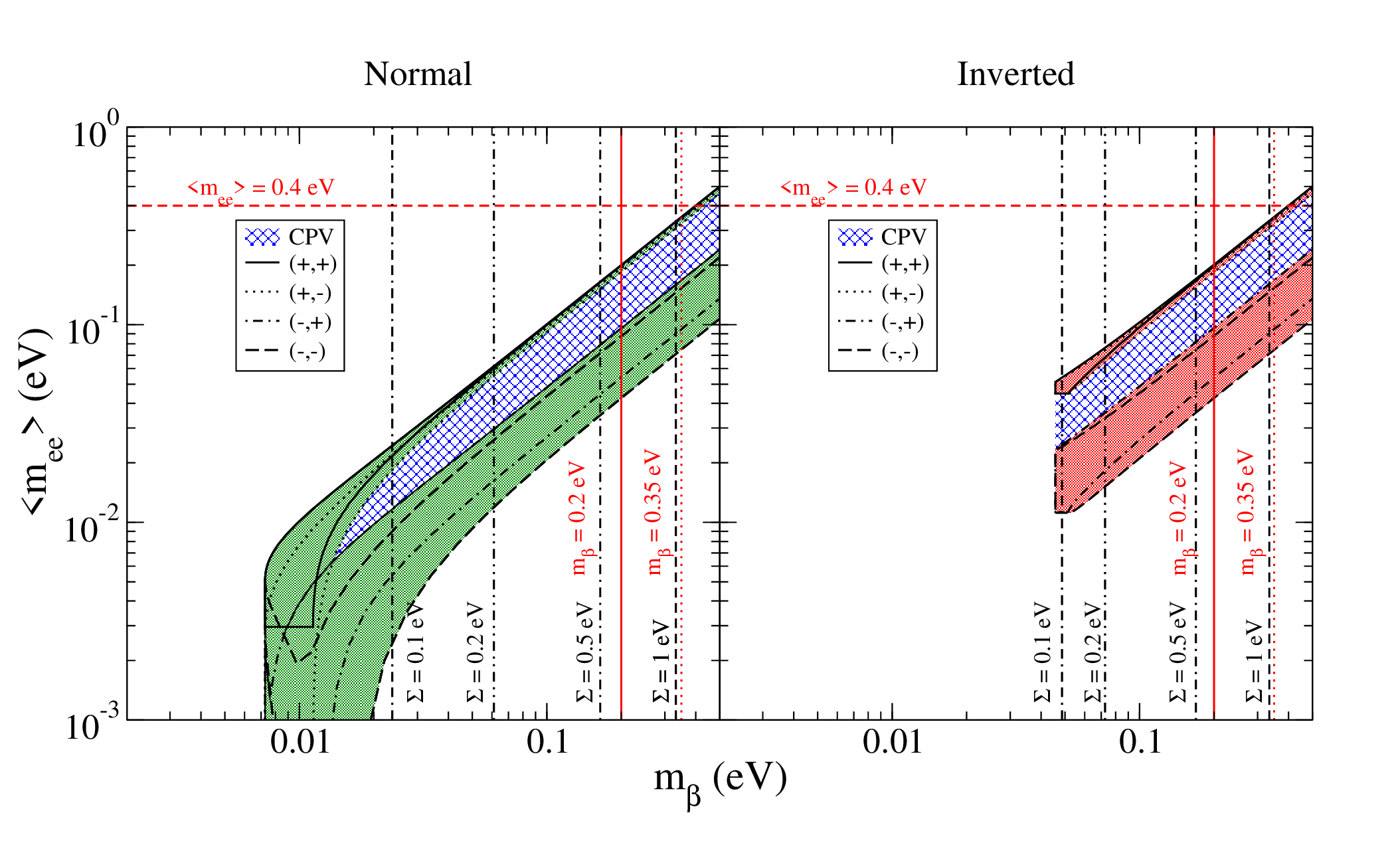}
\end{center}
\vspace*{8pt}
\caption{\label{fig:meff_obs}Effective mass against sum of masses $\Sigma$ and kinematic
neutrino mass $m_\beta$ for the $3\sigma$ ranges of the oscillation
parameters. The green and red shaded areas are the
general $3\sigma$ ranges, while the blue shaded areas can only be
realized if the CP phases take non-trivial values.  
Prospective future values of $\Sigma$ and $m_\beta$ are also given.}
\end{figure}

\begin{figure}[t]
\begin{center}
\input{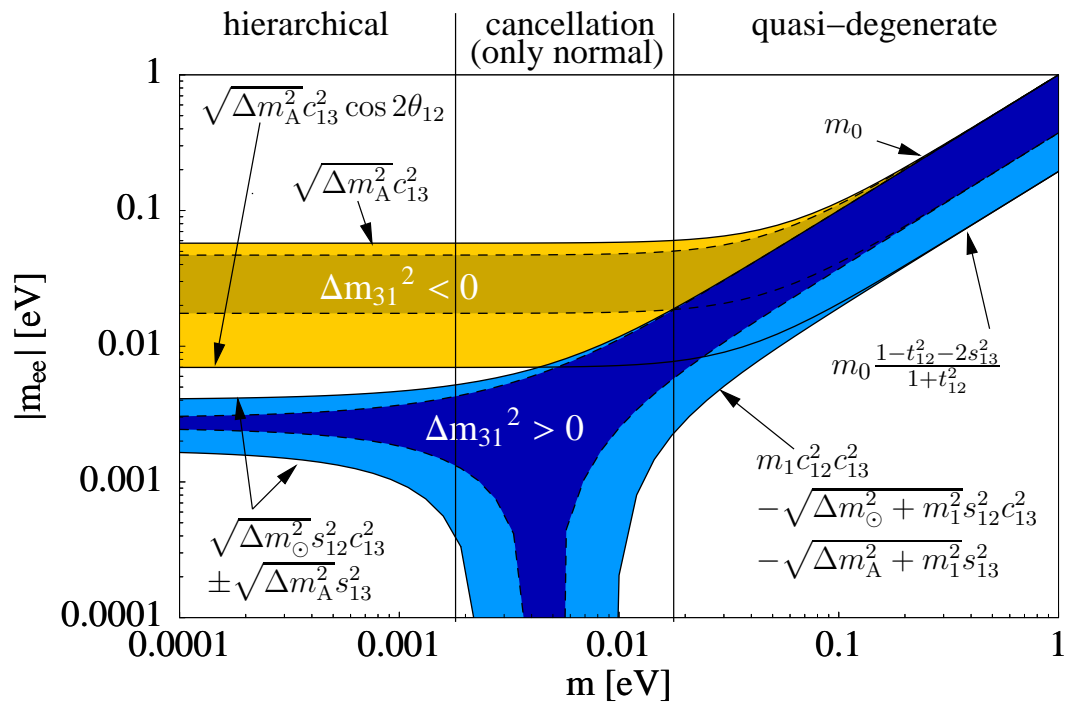}
\vspace*{8pt}
\caption{\label{fig:lovely_isn't_it?}The main properties 
of the effective mass as function of the smallest neutrino mass. Here
$m_0$ denotes the common mass for quasi-degenerate neutrinos and
$t_{12} = \tan \theta_{12}$. 
We indicate the relevant formulae and the three important regimes: 
hierarchical, cancellation and 
quasi-degeneracy. See \protect \cite{Lindner:2005kr}.}
\end{center}
\end{figure}

In the standard and symmetrical parametrization of the PMNS matrix we have 
\begin{eqnarray}
|m_{ee}^{(1)}| &=& m_1 \, |U_{e1}|^{2} = m_1 \, c_{12}^{2} \, c_{13}^{2} \,,
\nonumber\\
|m_{ee}^{(2)}| &=& m_2 \, |U_{e2}|^{2} = m_2 \, s_{12}^{2} \, c_{13}^{2} \,,\\
|m_{ee}^{(3)}| &=& m_3 \, |U_{e3}|^{2} = m_3 \, s_{13}^{2} \, .\nonumber
\end{eqnarray}
The individual masses can, using eq.~(\ref{eq:masses}), be expressed in
terms of the smallest mass and the mass-squared differences, whose
currently allowed ranges, as well as those of the mixing angles,
are given in table \ref{tab:osc}. 
A typical analysis of the effective mass would plot it against the smallest
neutrino mass \cite{Vissani:1999tu,Bilenky:1999wz,Bilenky:2001rz,Czakon:2001uh,Pascoli:2002xq,Feruglio:2002af,Pascoli:2003ke,Lindner:2005kr,Fogli:2006yq,Pascoli:2007qh,Fogli:2008ig}, while varying the Majorana phases and/or the
oscillation parameters. This results in figure \ref{fig:meff_mass}, for
which the best-fit values and 3$\sigma$ ranges of the oscillation parameters have been
used. The blue shaded areas are interesting because they can only be
covered if the CP phases are non-trivial (see section \ref{sec:CP}). It is also interesting to
plot the effective mass against the other neutrino mass observables  
$\Sigma$ and $m_\beta$, which is shown in figure
\ref{fig:meff_obs}. Such plots are helpful if more than one of the
complementary neutrino mass experiments finds a signal. 
The simple analytical expressions for \meff~in certain extreme
cases are given in figure \ref{fig:lovely_isn't_it?}.

\begin{table}[ht]
\caption{\label{tab:exp_res}Life-time limits on \onbb~at 90 \% C.L.~and limits on the effective mass
extracted with the matrix element compilation from figure
\ref{fig:nme}. Note that the range of nuclear
matrix elements leads to a range for the upper limits on \meff. 
Phase space factors are taken from
\cite{Suhonen:1998ck}.}
\begin{tabular}{@{}cccccc@{}} \br
Isotope & $T_{1/2}^{0\nu}$ [yrs]  & Experiment  & $G$ [10$^{-14}$ yrs$^{-1}$] 
& $\meff^{\rm lim}_{\rm min}$ [eV] &
$\meff^{\rm lim}_{\rm max}$ [eV]  \\ \hline 
$^{48}$Ca & $5.8 \times 10^{22}$ & CANDLES  \cite{Umehara:2008ru} &
6.35 & 3.55  &   9.91\\ \hline
$^{76}$Ge & $1.9 \times 10^{25} $ & HDM
\cite{KlapdorKleingrothaus:2000sn} & 0.623 &  0.21  &   0.53\\ 
 & $1.6 \times 10^{25} $ & IGEX  \cite{Aalseth:2002rf} &  & 0.25    &  0.63 \\ \hline
$^{82}$Se & $3.2 \times 10^{23} $ & NEMO-3  \cite{Arnold:2005rz} &
2.70 & 0.85 &   2.08 \\ \hline
$^{96}$Zr & $9.2 \times 10^{21} $ & NEMO-3  \cite{Argyriades:2009ph}&
5.63 &   3.97  &  14.39\\ \hline
$^{100}$Mo & $1.0 \times 10^{24} $& NEMO-3  \cite{Arnold:2005rz} &
4.36 &  0.31  &   0.79\\ \hline
$^{116}$Cd & $1.7 \times 10^{23} $ & SOLOTVINO
\cite{Danevich:2003ef}& 4.62 &  1.22  &  2.30\\ \hline
$^{130}$Te &  $2.8 \times 10^{24} $ &
CUORICINO  \cite{Arnaboldi:2005cg} & 4.09 & 0.27 & 0.57 \\ \hline
$^{136}$Xe &  $1.6 \times 10^{25} $& EXO-200
\cite{Auger:2012ar}&   4.31 & 0.15  &   0.36\\
 &  $5.7 \times 10^{24} $& KamLAND-Zen
\cite{KamLANDZen:2012aa} &    & 0.25   & 0.60  \\ \hline
$^{150}$Nd & $1.8 \times 10^{22} $ & NEMO-3  \cite{Argyriades:2008pr}
& 19.2 &   2.35  &   5.08\\ \br 
\end{tabular}\\[.6cm]
%\end{table}
%
%\begin{table}[hb]
\caption{\label{tab:future}Details of the most advanced
experiments. Given are life-time sensitivity and the expected limit on
\meff, using the NME compilation from figure \ref{fig:nme}. Note that the range of nuclear
matrix elements leads to a range for the expected sensitivity on \meff. }
%\begin{indented}
\begin{tabular}{ccccccc} 
\hline
Experiment & Isotope & Mass [kg] & Sensitivity  
%Sensitivity 
& Status & Start of  & Sensitivity \\
& &  & $T_{1/2}^{0\nu}$ [yrs] & 
& 
data-taking & $\langle m_{\nu} \rangle$ [eV]\\ \br 
GERDA & $^{76}$Ge & 18 & $3\times10^{25}$ 
%70-200 
& running & $\sim$ 2011 & 0.17-0.42\\
  &  & 40 & $2\times10^{26}$ 
%70-200 
& construction & $\sim$ 2012 & 0.06-0.16\\ 
& & 1000 & $6\times10^{27}$  
%10-40 
& R\&D & $\sim$ 2015& {0.012-0.030}\\ \hline
CUORE & $^{130}$Te & 200 & $6.5\times10^{26}$$^{*}$ 
%20-50 
& construction & $\sim$ 2013 & { 0.018-0.037}\\ 
& & & $2.1\times10^{26}$$^{**}$ & 
%40-90 
& & {0.03-0.066}\\ \hline
MAJORANA & $^{76}$Ge & 30-60 & $(1-2)\times10^{26}$ 
%70-200 
& construction & $\sim$ 2013 & 0.06-0.16\\
& & 1000 & $6\times10^{27}$  
%10-40 
& R\&D & $\sim$ 2015&  {0.012-0.030}\\ \hline
EXO 
& $^{136}$Xe & 200 & $6.4\times10^{25}$ 
%100-200 
& running & $\sim$ 2011 & 0.073-0.18\\
& & 1000 & $8\times10^{26}$  
%30-60 
& R\&D & $\sim$ 2015& 0.02-0.05\\ \hline
SuperNEMO & $^{82}$Se & 100-200 & $(1-2)\times10^{26}$ 
%40-100 
& R\&D & $\sim$ 2013-15& 0.04-0.096\\
\hline
KamLAND-Zen & $^{136}$Xe & 400 & $4\times10^{26}$ 
%40-80 
& running & $\sim$ 2011 & 0.03-0.07\\
& & 1000 & $10^{27}$ 
%25-50 
& R\&D & $\sim$ 2013-15& { 0.02-0.046}\\ \hline
SNO+ & $^{150}$Nd & 56 & $4.5\times10^{24}$ 
%100-300 
& construction & $\sim$ 2012 & 0.15-0.32\\
& &  500 & $3\times10^{25}$ 
%40-120 
&        R\&D  & $\sim$ 2015& 0.06-0.12\\ 
\br
\end{tabular}
%\end{indented}
\end{table}

\begin{figure}[b]
\begin{center}
\includegraphics[width=9cm,height=7.5cm]{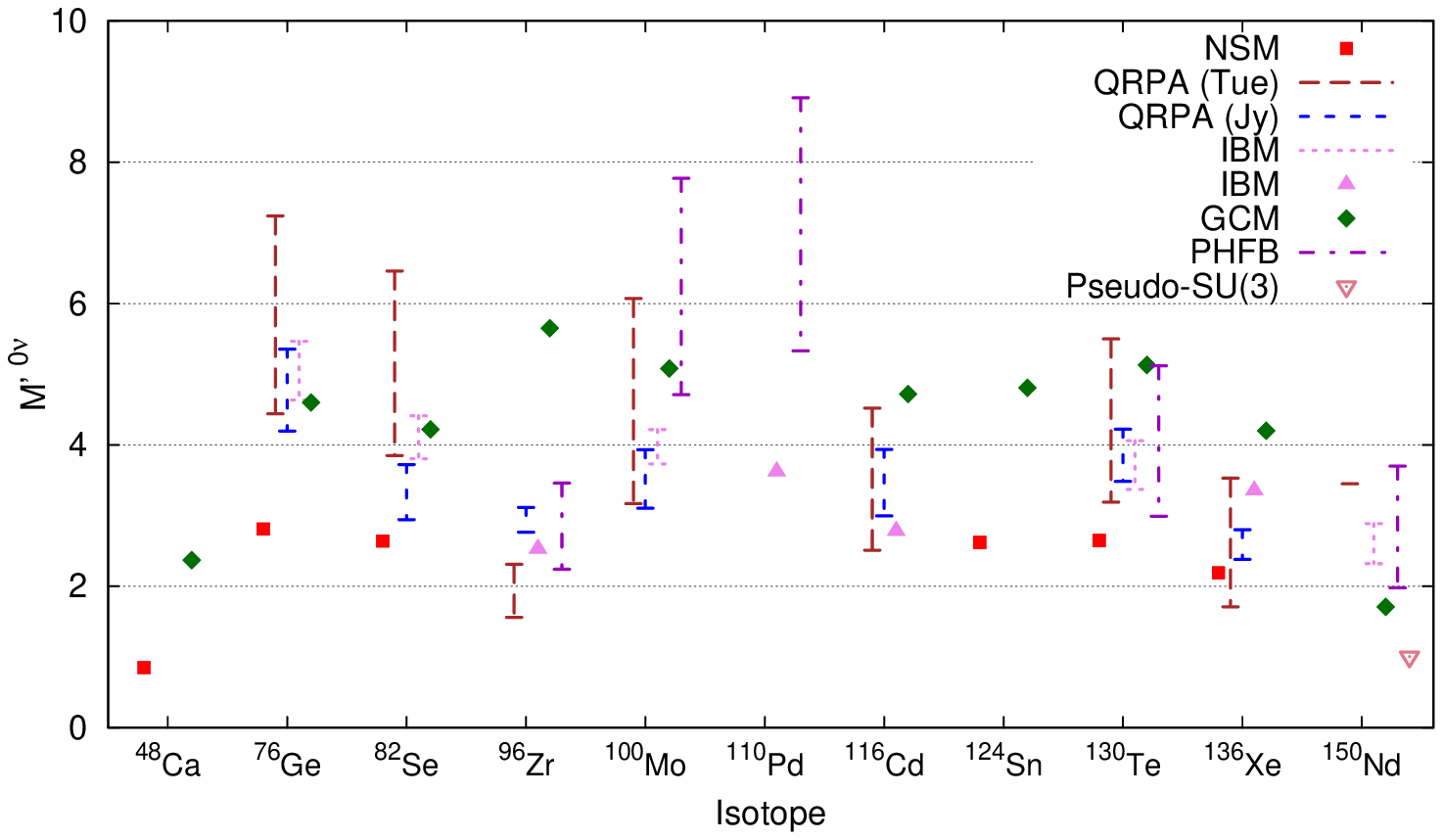}
\end{center}
\vspace*{8pt}
\caption{\label{fig:nme}Nuclear matrix element compilation for \obb, different
isotopes and calculational approaches. }
\end{figure}

%\newpage

In order to translate the effective mass into life-time, one needs to
give the phase space factor $G(Q,Z)$ and the (range of) nuclear matrix
elements. It is an ongoing discussion on which NMEs one should use, and
what their uncertainty is. The contributions to the focus issue by
Faessler, Simkovic and Rodin, by Suhonen and Civitarese, by Menendez,
by Engel and by Vogel will discuss this at length. We will stick here
for definiteness to the compilation from \cite{Dueck:2011hu},
displayed in figure \ref{fig:nme} (see also recent compilations in 
\cite{GomezCadenas:2011it,Bilenky:2011tr,Faessler:2012ku}). The
current limits on \obb~and the required phase
space factors are given in table \ref{tab:exp_res}, together with the
resulting limits on the effective mass. Note that the range of nuclear
matrix elements leads to a range for the upper limits on \meff. The current limit on the
effective mass is therefore\footnote{There is a much debated claim
\cite{KlapdorKleingrothaus:2001ke} of observation, which corresponds
to a value of \meff~of about 0.2 to 0.6 eV.}  
\be \label{eq:limit_meff}
\meff \ls 0.4 ~{\rm eV} \, . 
\ee
We stress here that the very recent result from EXO-200
\cite{Auger:2012ar} has finally improved the long-standing best limit
from the Heidelberg-Moscow experiment
\cite{KlapdorKleingrothaus:2000sn}. 
Details of the most
advanced experiments in what regards their expected limits on the
effective mass can be found in table \ref{tab:future}. 
It is worth to take a closer look at the interesting features of the
effective mass shown in figure \ref{fig:lovely_isn't_it?}: 
\begin{itemize}
\item in the normal hierarchy case, the effective mass is smallest,
somewhere in the meV regime: 
\be \label{eq:meff_NH_scale}
\meff^{\rm NH} \sim 
\sqrt{\dms} \, \sin^2 \theta_{12} +\sqrt{\dma} \, \sin^2 \theta_{13}
 \simeq 0.004 \, \rm eV \, .
\ee
The meV scale of the effective mass should be the final goal of
experiments. The half-lifes corresponding to
meV effective masses are $10^{28}$ to $10^{29}$ yrs; 
\item for certain values of the parameters the effective mass can even
vanish. This unfortunate situation will be discussed in section
\ref{sec:0}; 
\item for the inverted hierarchy the effective mass cannot vanish. 
In the limit of negligible $m_3 \, |U_{e3}|^2$ one has 
\be
\hspace{-2cm}\meff^{\rm IH}_{\rm max} \equiv \sqrt{\dma}\,c_{13}^{2} \le 
\meff^{\rm IH} \le \sqrt{\dma}\,c_{13}^{2} \cos 2 \theta_{12} 
\equiv \meff^{\rm inv}_{\rm min} \, . 
\label{eq:meeIHleft}
\ee
Due to the non-maximal value of
$\theta_{12}$ the minimal value of the effective mass is
non-zero \cite{Pascoli:2002xq}. Therefore, if limits below the
minimal value 
\begin{equation}
\meff^{\rm{inv}}_{\rm{min}} = \meff^{\rm{IH}}_{\rm{min}} = 
\left(1 - |U_{e3}|^2 \right)  \sqrt{\dma} \left(1 - 2 \, \sin^2
\theta_{12} \right) , \label{eqn:meffminih}
\end{equation}
are reached by an experiment, the inverted mass ordering is ruled out
if neutrinos are Majorana particles. If we knew by independent
evidence that the mass ordering is inverted (by a long-baseline
oscillation experiment or a galactic supernova observation) then we
would rule out the Majorana nature of neutrinos. Of course, one has to
assume here that no other lepton number violating mechanism
interferes. The two scales of
\meff~corresponding to the minimal and maximal value in case
of the inverted hierarchy, given in eq.~(\ref{eq:meeIHleft}), should
be the intermediate or long-term goal 
of experiments.  
The typical effective mass values of order $0.02$
eV are one order of magnitude
larger than those for the normal hierarchy and
roughly one order of magnitude smaller than for quasi-degenerate
neutrinos. They correspond to half-lifes of order 
$10^{26}$ to $10^{27}$ yrs; 
\item in the quasi-degenerate regime, both mass orderings are
indistinguishable, the effective mass in this case reads 
\be
\meff^{\rm QD} = m_0  \left|  c_{12}^{2} \, 
c_{13}^{2} + s_{12}^{2} \, c_{13}^{2}\, 
e^{2i\alpha} + s_{13}^{2} \, e^{2i\beta}
\right| .
\ee
This expression cannot vanish either. It corresponds, for $\meff^{\rm QD}
\simeq 0.1$ eV, to half-lifes in the ballpark of $10^{25}$ to
$10^{26}$ yrs, hence the QD mass scheme is currently tested;  
\item it is evident from the approximate expressions of the effective
mass and figure \ref{fig:lovely_isn't_it?} that the recently found
sizable value of $U_{e3}$ is currently of minor importance for \obb. The 
difference between the maximal effective mass in the normal hierarchy 
and the minimal value in the inverted hierarchy shrinks a bit \cite{Lindner:2005kr}. In
addition, the area in which the effective mass vanishes becomes
larger. These properties pose no problem for  current and
next-generation experiments. 

\end{itemize}
The approximate expressions for the effective mass and the other
neutrino mass observables are summarized in table \ref{tab:mass_obs}. Table \ref{tab:mass_obs_compl} attempts to illustrate the
complementarity of neutrino mass observables. Prospective sensitivity
values of $m_\beta = 0.2$ eV, $\meff = 0.02$ eV, and $\Sigma = 0.1$ eV
are assumed and the interpretation of positive and/or negative results
in all 3 approaches is given. 

\begin{table}[t]
\caption{\label{tab:mass_obs}Approximate analytical expressions for the neutrino mass
observables for the extreme cases of the mass ordering. For \obb~the
typical (isotope-dependent) half-lifes are also given.}
\begin{center}
\begin{indented}
\item[]\begin{tabular}{cccc}
 &  $\Sigma$ &  $m_\beta$ 
&  \meff \\ \br
 NH &   $\sqrt{\dma}$ 
&   $\sqrt{\sss \dms + |U_{e3}|^2  \dma}  $ 
&  $\left| \sss  \sqrt{\dms} + |U_{e3}|^2 \sqrt{\dma} 
e^{2 i (\alpha - \beta)} \right| $ \\ 
& $\simeq 0.05$ eV & $\simeq 0.01$ eV & $\sim 0.004$ eV $\Rightarrow
T^{0\nu}_{1/2} \gs 10^{28 - 29}$ yrs \\ \hline 
 IH &   $2  \sqrt{\dma} $  
&   $\sqrt{\dma} $ &   $\sqrt{\dma}  
\sqrt{1 - \sin^2 2 \theta_{12} \sin^2 \alpha}$  \\ 
& $\simeq 0.1$ eV & $\simeq 0.05$ eV & $\sim 0.02$ eV $\Rightarrow T^{0\nu}_{1/2} \gs 10^{26 - 27}$ yrs\\
\hline 
 QD &   $3  m_0$  &   $m_0$ 
&   $m_0 \sqrt{1 - \sin^2 2 \theta_{12}  
\sin^2 \alpha}$  \\ 
& & & $\gs 0.1$ eV $\Rightarrow T^{0\nu}_{1/2} \gs 10^{25 - 26}$ yrs\\ \br
\end{tabular} 
\end{indented}
\end{center}
\end{table}

{\scriptsize 
\begin{table}[t]
\caption{\label{tab:mass_obs_compl}``Neutrino mass matrix'' for the
present decade. It is assumed that KATRIN will reach its sensitivity
limit of $m_\beta = 0.2$ eV, that \obb-experiments can obtain values
down to $\meff = 0.02$ eV, and that cosmology can probe the sum of
masses down to $\Sigma = 0.1$ eV. With ``yes'' a positive measurement
is denoted, while ``no'' refers to no observation at the limit. 
M denotes Majorana neutrino, D Dirac
neutrino, N-SI non-standard interpretation of \obb, N-SC non-standard cosmology. }
%\begin{indented}
\hspace{-1.8cm}\begin{tabular}{cccccccc}
& & \multicolumn{2}{c}{ KATRIN} & \multicolumn{2}{c}{${\hspace{-.3cm}0\nu\beta\beta}$} &
\multicolumn{2}{c}{\hspace{.6cm} cosmology} \\ \hline 
& & yes & no & yes & no & yes & no \\ \br 
 KATRIN & $\ba \mbox{yes} \\ \mbox{no} \ea $ & $ \ba-\\- 
\ea  $ & $ \ba-\\-\ea  $ & $ \ba \mbox{QD + M}
\\ \mbox{N-SI} \ea $  & $ \ba \mbox{QD + D} \\ \mbox{low IH or NH
or D} \ea $ & $\ba \mbox{QD} \\ m_\nu \ls 0.1 \, \mbox{eV or N-SC}
\ea $ & $\ba \mbox{N-SC} \\ \mbox{NH} \ea $ \\ \hline
 
${0\nu\beta\beta}$ & $\ba \mbox{yes} \\ \mbox{no} \ea $ 
& $ \ba \bullet \\ \bullet \ea  $ & $  \ba \bullet \\ \bullet \ea  $
& $ \ba-\\- 
\ea  $ & $ \ba-\\-\ea  $  
& $\ba \mbox{(IH or QD) + M} \\ \mbox{low IH or (QD + D)} \ea $ 
& $\ba \mbox{N-SC or N-SI} \\ \mbox{NH} \ea $ \\ \hline 

 cosmology & $\ba \mbox{yes} \\ \mbox{no} \ea $ & $ \ba \bullet \\
\bullet \ea  $ & $ \ba \bullet \\ \bullet \ea  $ & $ \ba \bullet \\
\bullet \ea  $ & $ \ba \bullet \\ \bullet \ea  $ & $ \ba-\\
- \ea  $ & $ \ba- 
\\-\ea  $ \\ \br
\end{tabular} 
%\end{indented}
\end{table}
}

\begin{figure}[t]
\begin{center}
\includegraphics[height=5cm,width=6cm]{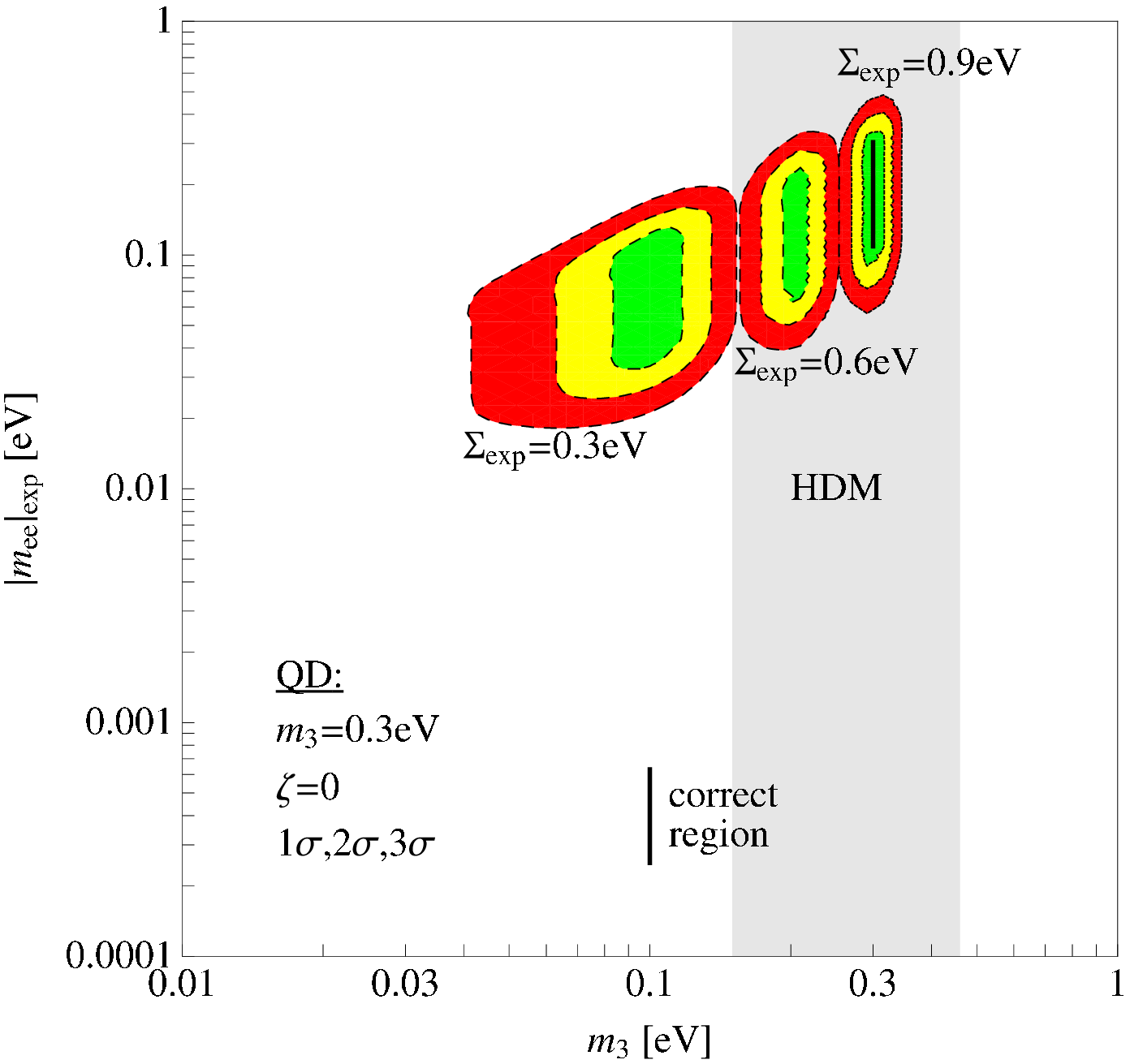}
\includegraphics[height=5cm,width=6cm]{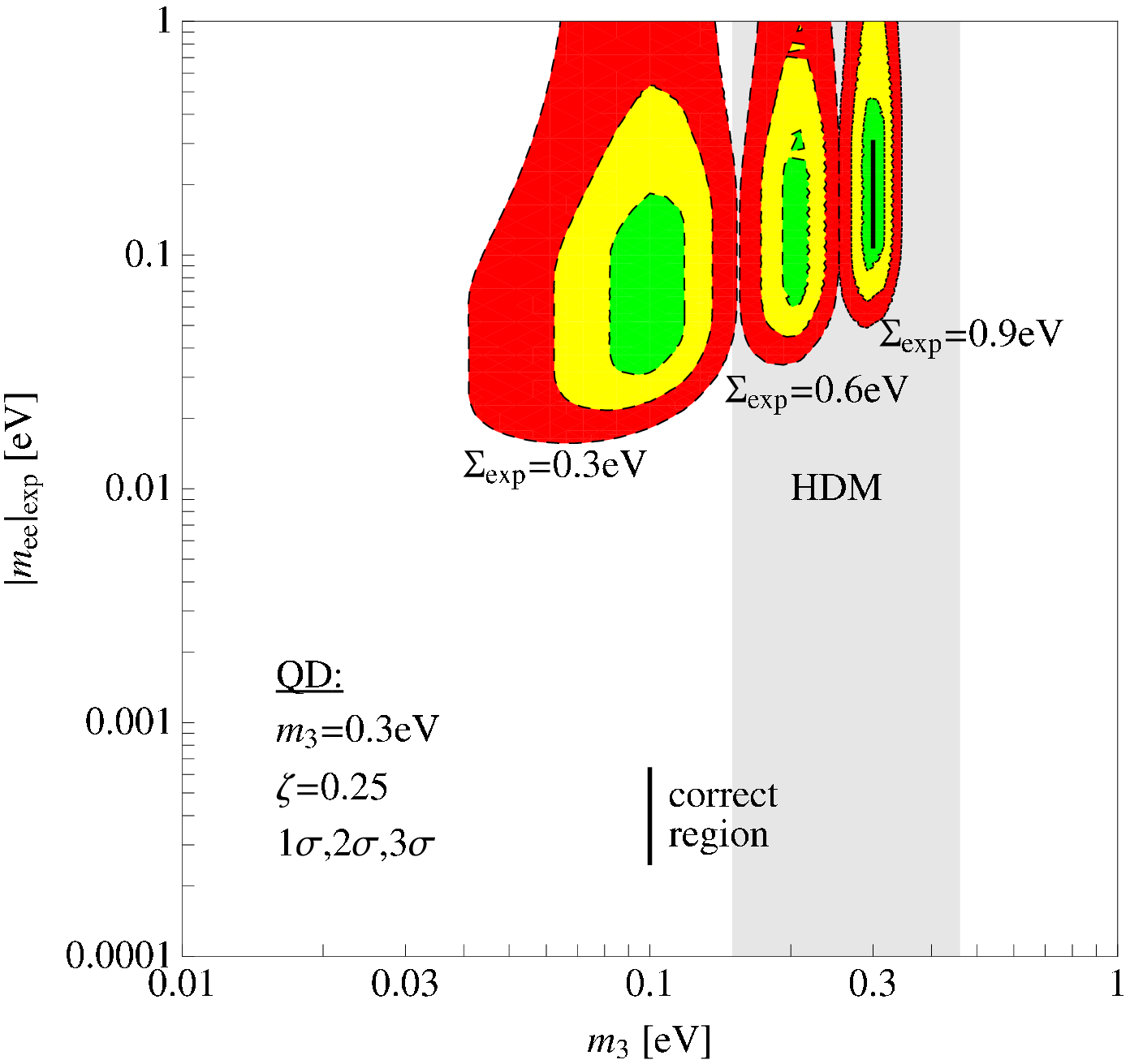}
\end{center}
\vspace*{8pt}
\caption{\label{fig:MMR}1$\sigma$, 2$\sigma$ and 
3$\sigma$ regions in the $m_3$-$\meff_{\rm exp}$ plane for a
quasi-degenerate neutrino mass scenario. The left plot assumes no 
NME uncertainty, the right plot assumes 25\% uncertainty. The correct
(solid line) as well as two possible incorrect cosmological 
measurements (dashed lines) are used. See \protect \cite{Maneschg:2008sf}.}
\end{figure}

\subsection{\label{sec:mnu}Neutrino mass}

With a current limit $\meff_{\max}^{\rm exp}$ on the effective mass,
which corresponds to the quasi-degenerate regime, one can obtain the
following limit on the neutrino mass: 
\be \label{eq:m0_lim}
m_0 \le  \meff_{\max}^{\rm exp} \,  \frac{1 + \tan^2 \theta_{12}}
{1 - \tan^2 \theta_{12} - 2 \, |U_{e3}|^2 } 
\equiv \meff_{\max}^{\rm exp} \,  f(\theta_{12}, \theta_{13}) \, .
\ee
Note that $m_0$ is in the QD regime identical to $m_\beta$, i.e.~to
the quantity measured in Kurie-plot experiments.  
The function $f(\theta_{12}, \theta_{13})$ varies from 2.75 to 3.43
at $1\sigma$ and from 2.28 to 4.66 at $3\sigma$. The limit on the
effective mass is about 0.4 eV (see (\ref{eq:limit_meff})), and hence
$m_0 \le 1.4$ eV and 1.9 eV, respectively. Therefore, the current limit
on $m_0$ from \obb~is very similar to the one from the Mainz
experiment. In the QD regime one further has the relation $m_0 =
\Sigma/3$.

Perhaps more interesting is the determination of the neutrino mass
scale in future experiments if information from complementary 
neutrino mass observables is combined. For instance,
Ref.~\cite{Maneschg:2008sf} (see also \cite{Pascoli:2005zb}) has 
performed a statistical analysis of prospective data, see figure \ref{fig:MMR}. Using
realistic errors for the experimental quantities, it was found that
the neutrino mass can be determined (at $3\sigma$) with 15 \% to 25 \%
uncertainty, depending on the NME uncertainty. The precision is largely determined by
cosmology, the uncertainty of the oscillation parameters is of little
importance.

\begin{figure}[th]
\begin{center}
\includegraphics[width=7cm,height=6cm]{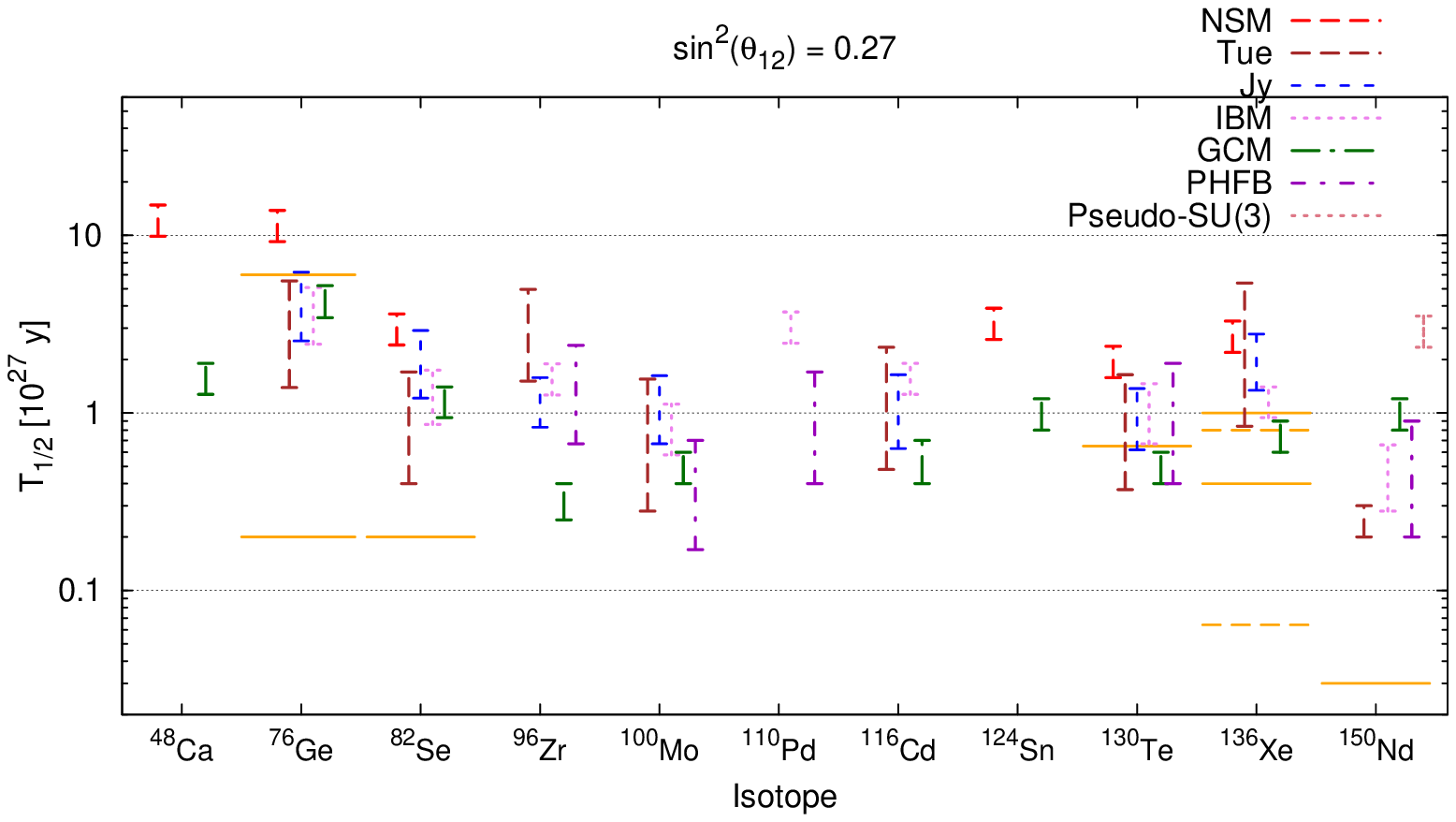} \quad
\includegraphics[width=7cm,height=6cm]{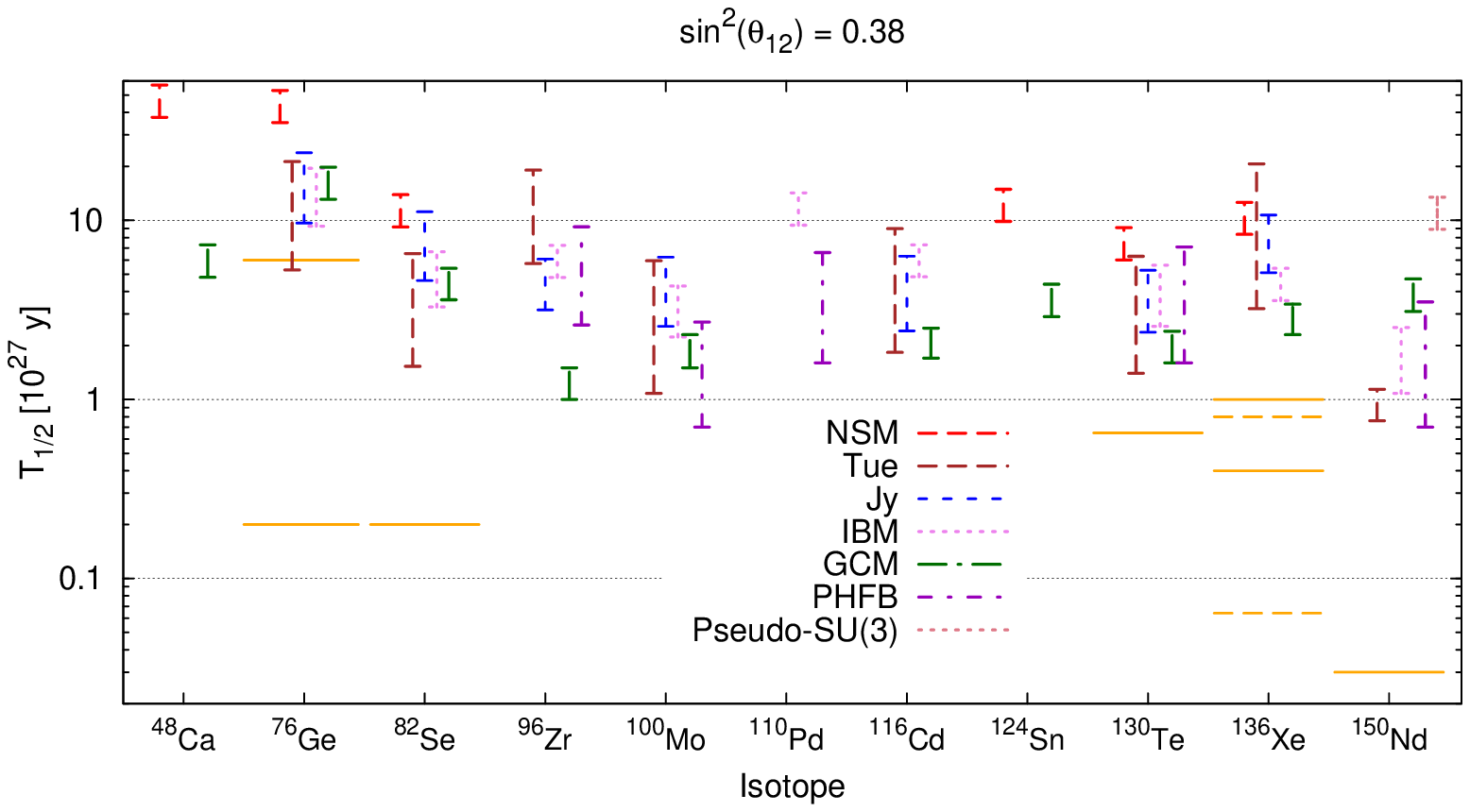}

\includegraphics[width=7cm,height=6cm]{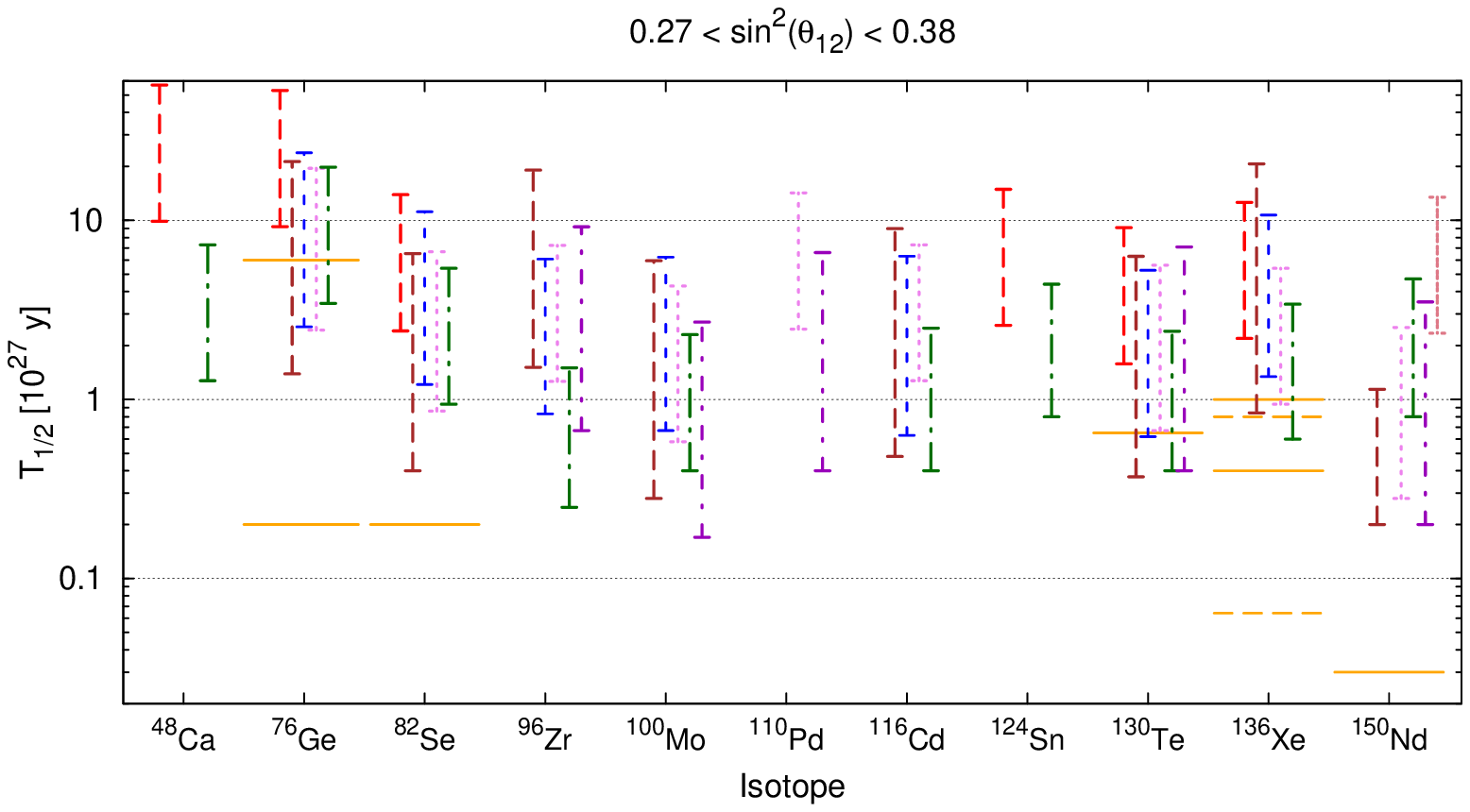} 
\includegraphics[width=7cm,height=6cm]{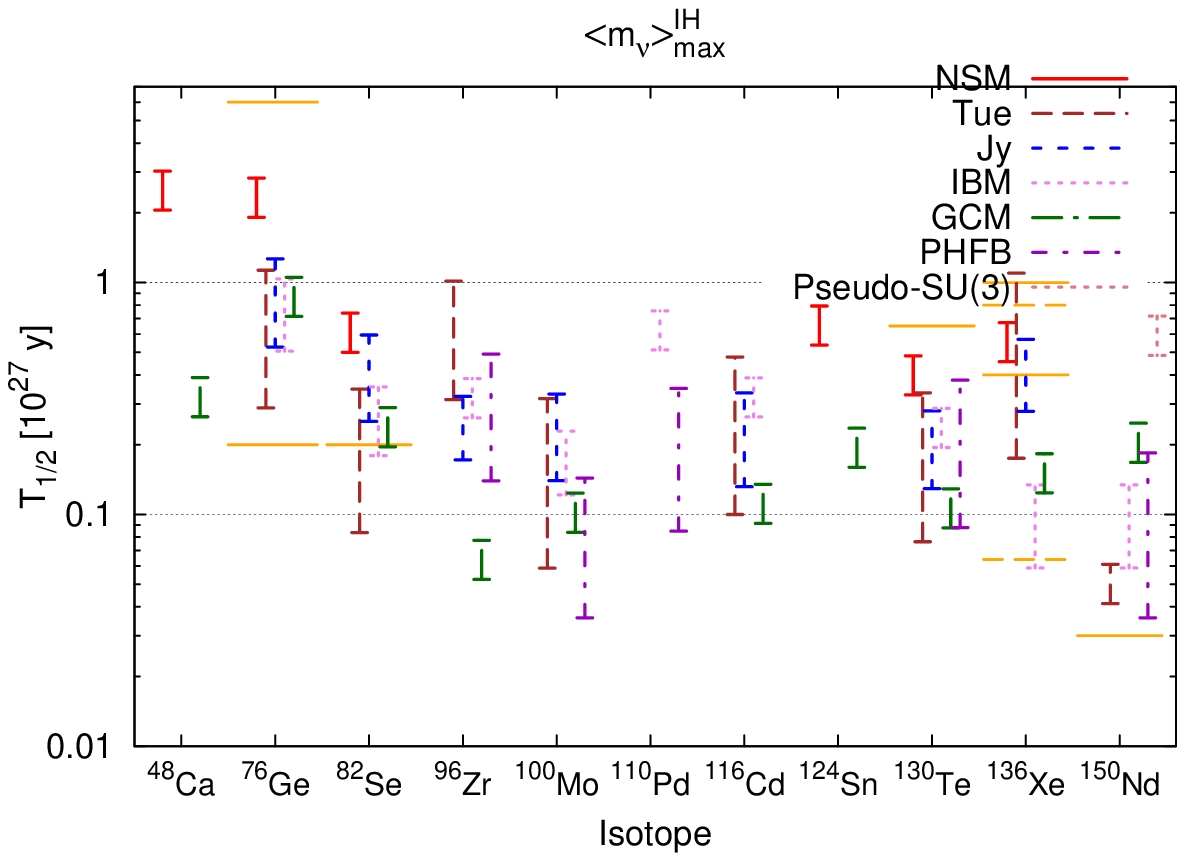} 

\end{center}
\vspace*{8pt}
\caption{\label{fig:isohl}Required half-life sensitivities to exclude
and touch the inverted hierarchy for different values of
$\theta_{12}$, using the compilation of NMEs from figure
\ref{fig:nme}. The upper plots show the necessary half-lifes for $\sin^2 
\theta_{12} = 0.27$ (upper left) and $\sin^2 \theta_{12} = 0.38$
(upper right). The lower left plot includes the current $3\sigma$
uncertainty for $\theta_{12}$. The lower right plot shows the
necessary half-lifes in order to touch the inverted ordering, which is
independent on $\theta_{12}$.  The small horizontal lines show
expected half-life sensitivities at 90\% C.L.~of running and planned
\obb-experiments. When two sensitivity expectations are
given for one experiment they correspond to near and far time
goals. See \protect \cite{Dueck:2011hu}.}
\end{figure}

\subsection{\label{sec:NHIH}Neutrino mass ordering}

From figure \ref{fig:meff_mass} the interesting possibility of ruling out
the inverted mass ordering becomes obvious. 
The minimal value of the effective mass, repeated here for
convenience, is non-zero and given by  
\be
\meff^{\rm{inv}}_{\rm{min}} = 
\left(1 - |U_{e3}|^2 \right)  \sqrt{\dma} \left(1 - 2 \, \sin^2
\theta_{12} \right) . 
\ee
If a limit on the effective mass below this value is obtained, the
inverted ordering is ruled out if neutrinos are Majorana particles. In
case the mass ordering is known to be inverted 
then the Majorana nature of neutrinos would be ruled out.

One can translate the effective mass necessary to rule out (or touch)
the inverted hierarchy into half-lifes, see
table \ref{tab:future} and figure \ref{fig:isohl}.  We note that currently running
experiments will not fully probe the inverted hierarchy regime. 
The crucial dependence on
$\theta_{12}$ has recently been discussed in \cite{Dueck:2011hu}. 
The current $3\sigma$ range of
$\theta_{12}$ corresponds to an uncertainty of a factor $\simeq 1.77$ in the minimal value of
the effective mass, which is of the same order as the current uncertainty in
the NMEs.  
The factor $1.77$ due to $\theta_{12}$ corresponds to a factor of $1.77^2 \simeq 3.13$ in half-life. 
In experiments with background, see (\ref{eq:Texp}), this means a 
rather non-trivial combined factor of $1.77^4 \simeq 10$ in the product of 
measuring time, energy resolution, background index and detector
mass. Therefore, a precision determination of the
solar neutrino mixing angle would be very desirable to
evaluate the requirements and physics potential of upcoming
\obb-experiments in order to test the inverted ordering
\cite{Dueck:2011hu}.

\subsection{\label{sec:CP}Majorana phases}

Determining a Majorana CP phase from \onbb~is probably the most 
difficult physics goal related to \obb~\cite{Fukuyama:1997ez,Matsuda:2000jx,Rodejohann:2000ne,Matsuda:2000iw,Pascoli:2001by,Rodejohann:2002ng,Nunokawa:2002iv}. Note that 
there are two Majorana phases and only one observable, \meff, and thus
only one (or a combination) of the phases can be
extracted in principle. In addition, complementary information on the neutrino
mass scale has to be put in for such a measurement, and only for the 
inverted ordering or the quasi-degenerate scheme a phase determination
is conceivable at all. 

\begin{figure}[t]
\begin{center}
\includegraphics[height=7.8cm,width=12cm]{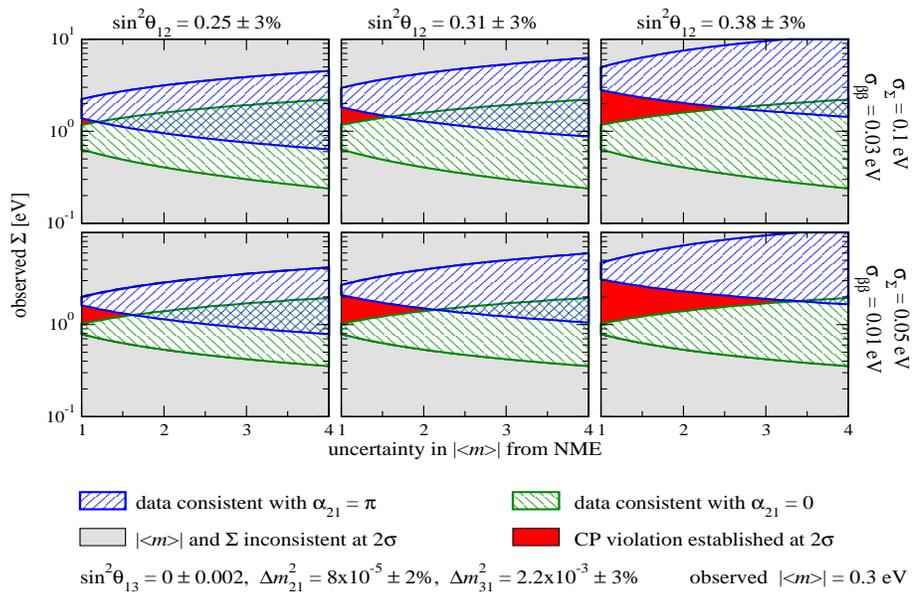}
\end{center}
\vspace*{8pt}
\caption{\label{fig:2005zbCP}
Constraints on the Majorana phase $2\alpha$ (here called $\alpha_{21}$) at 95\%~C.L.\
   from an observed $\meff_\mathrm{exp} = 0.3$~eV and prospective 
   data on $\Sigma$, as a function of the NME uncertainty factor. 
Shown are the regions in which the
   data are consistent with a CP conserving value 
    (hatched), observed $\Sigma$ is
   inconsistent with $\meff_\mathrm{exp}$ (light-shaded), and 
   Majorana CP violation is established (red/dark-shaded).
Taken from \protect \cite{Pascoli:2005zb}.}
\end{figure}

A somewhat pessimistic point of view was presented in
\cite{Barger:2002vy}, whereas \cite{Pascoli:2002qm} found the
requirements not too unrealistic.  Recalling figure
\ref{fig:meff_obs}, it is clear that in experiments one should 
find results lying in the areas indicated with 
``CPV'', which are however smeared by experimental and theoretical
uncertainties.  Neglecting
$\theta_{13}$, the effective mass for the IH or QD cases is proportional to 
\be
\meff \propto \left|\cos^2 \theta_{12} + e^{2 i \alpha} \, \sin^2 
\theta_{12}  \right| = \sqrt{1 - \sin^2 2 \theta_{12} \, \sin^2 \alpha} \, . 
\ee
Therefore, the larger $\theta_{12}$ is, the more promising it is to
extract $\alpha$ from measurements. Recall that ruling out the inverted
mass ordering is easier if $\theta_{12}$ is small. A detailed statistical analysis has been performed
in \cite{Pascoli:2005zb}, see figure \ref{fig:2005zbCP}. 
One can see that, as expected, for larger values of
$\theta_{12}$ the areas in parameter space become larger. 

%For instance, if $\sin^2\theta_{12} \gs 0.3$ 
%and $\simeq 10\%$ errors in the measured $\meff_\mathrm{exp}$ and
%$\Sigma$ are present, the NME has to been known to better than within a factor of 1.5.
%For smaller values of the errors, $\sigma_{\beta\beta} \simeq 0.01$~eV
%and $\sigma_\Sigma \simeq 0.05$~eV, Majorana CP-violation could be 
%established for larger NME uncertainty. Finally, the 
%Majorana phase $2\alpha$ has to have a value approximately 
%in the interval $\sim (\pi/4 - 3\pi/4)$. In the inverted hierarchy the
%required errors have to be smaller, and the determination of the
%phase is more challenging. 

\subsection{\label{sec:0}Vanishing effective mass}

Unfortunately, the normal mass hierarchy can allow for 
complete cancellation of the effective mass. In terms of 
figure \ref{fig:FD_mass_mech}, this ``cancellation regime'' means that a triangle
can be formed. Neglecting $m_1$, one needs
\be
\frac{m_2}{m_3} = \frac{\tan^2 \theta_{13}}{\sin^2 \theta_{12}} \simeq
3 \, \tan^2 \theta_{13} \simeq 0.08 \, . 
\ee
The Majorana phases need to be such that the two surviving terms have
opposite sign. For the general case one finds \cite{Dev:2006if}
\bea \D
\hspace{-3cm}\cos 2\alpha =\frac{
m_{3}^{2} \,s_{13}^{4}-c_{13}^{4}( m_{1}^{2} \, c_{12}^{4}+m_{2}^{2}
\, s_{12}^{4})}{
2 \,m_{1} \,m_{2} \,s_{12}^{2} \,c_{12}^{2} \,c_{13}^{4}} ~,~~ %\\ \D
\cos 2\beta =-\frac{
m_{3}^{2} \,s_{13}^{4}+c_{13}^{4} \,(m_{2}^{2} \,s_{12}^{4}-m_{1}^{2} \,c_{12}^{4})}{
2 \,m_{2} \,m_{3} \,s_{12}^{2} \,s_{13}^{2} \,c_{13}^{2}} \, . 
\eea
It may seem unnatural that the 7 parameters on which \meff~depends
conspire in such a way that the effective mass vanishes\footnote{See
\cite{Rodejohann:2011mu} for a list of corrections which can lead to
non-zero \meff. Note that the renormalization group running in the
effective theory is multiplicative and cannot generate a non-zero
value. Typically, if non-zero, the effective mass slightly increases from
low to high scale.}. However, 
recall that the effective mass is the $ee$ element of the Majorana
neutrino mass matrix. This matrix is 
generated by the underlying theory of mass generation, and texture
zeros occur frequently in such flavor models (cf.~section \ref{sec:flav}), see
\cite{Grimus:2004hf} for a general analysis and \cite{Jenkins:2008ms} 
for symmetries leading to $\meff=0$.

Experimentally, the effective mass can be considered zero if it is 
below $10^{-4}$ eV, no future experiment that currently is envisaged
can reach such low values. If one of the complementary neutrino mass
measurements finds a signal, then this means that neutrinos are QD or
IH, and therefore the effective mass cannot vanish.

\subsection{\label{sec:flav}Distinguishing flavor models}

There is an industry of model building in order to explain the
peculiar mixing structure of leptons that is so different to quark
mixing \cite{Altarelli:2010gt,Ishimori:2010au,Hirsch:2012ym}. 
Many models lead to the same neutrino mixing scheme, in particular
tri-bimaximal mixing (TBM): 
\[
\hspace{-2cm}U = \left(\bad
\sqrt{\frac 23} & \sqrt{\frac 13} & 0 \\
-\sqrt{\frac 16} & \sqrt{\frac 13} & \sqrt{\frac 12} \\
\sqrt{\frac 16} & -\sqrt{\frac 13} & \sqrt{\frac 12} 
\ea \right) \Rightarrow m_\nu = 
\left(
\bad 
A & B & B \\[0.2cm]
\cdot & \frac{1}{2} (A + B + D) & \frac{1}{2} (A + B - D)\\[0.2cm]
\cdot & \cdot & \frac{1}{2} (A + B + D)
\ea 
\right). 
\] 
The prediction $U_{e3}=0$ can be corrected to non-zero values by a
variety of mechanisms, and there are also many alternatives to TBM
\cite{Albright:2010ap}. 
It turns out that neutrino mass observables can help in disentangling
the vast amount of flavor symmetry models. One example is that the
flavor symmetry leads to correlations of the mass matrix elements,
which imply correlations of observables. For instance, the effective
mass could be correlated with the atmospheric neutrino parameter
$\sin^2 \theta_{23}$ \cite{Hirsch:2007kh}. Recall that in general $\theta_{23}$ has 
no influence on \meff. 

\begin{figure}
 
 \begin{center}
\includegraphics[height=8cm,width=12cm]{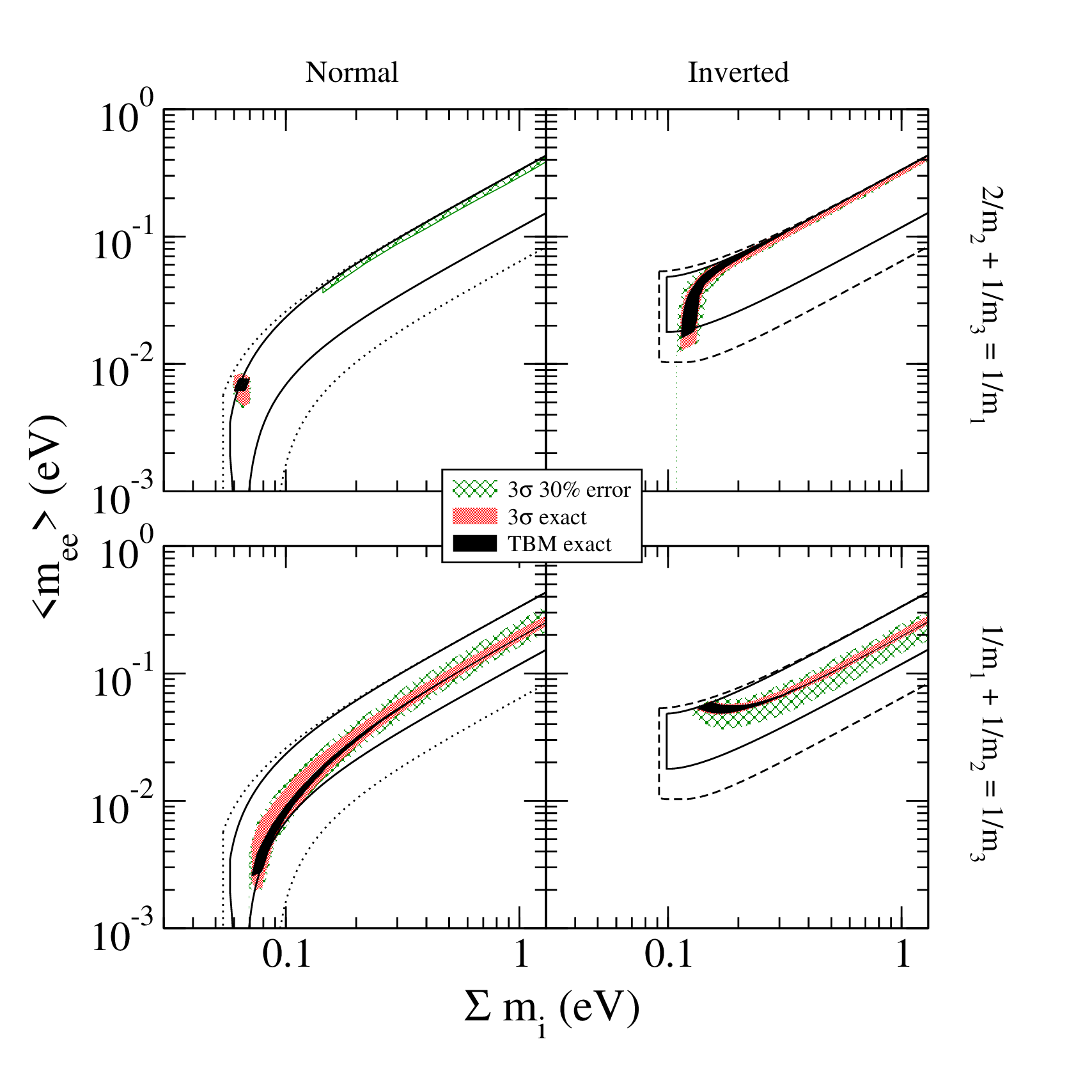}
\end{center}

\vspace*{8pt}
 \caption{Allowed regions in $\meff-\Sigma$ parameter space for the
sum-rules $\frac{2}{m_2}+\frac{1}{m_3}=\frac{1}{m_1}$ (top) and
$\frac{1}{m_1}+\frac{1}{m_2}=\frac{1}{m_3}$ (bottom), for both the
TBM (black)
and $3\sigma$ values (light red) of the oscillation data, as well as
for the sum-rules violated by 30\% (green hatches). See \protect \cite{Barry:2010yk}.
}
\label{fig:mass_flav}
\end{figure}

Another point are neutrino mass ``sum-rules'': 
the most general neutrino mass matrix giving rise to TBM is 
given above. 
As such, the (complex) eigenvalues $ A-B$, $A + 2 \, B$ and
$D$ are independent of the mixing angles: no matter what $A,B,D$
are, the PMNS mixing is given as above. However, very often the
structure of the mass matrix is simpler, and ``sum-rules'' between the neutrino masses
arise. Examples are $2/m_2 + 1/m_3 = 1/m_1$ or $1/m_2 + 1/m_3 =
1/m_1$ (here the masses are understood to be complex, i.e.~including
their Majorana phases). As a consequence of the sum-rules, not all
masses and phases are allowed, and detailed studies of the predictions can be found
in \cite{Barry:2010yk,Dorame:2011eb,Dorame:2012zv}. See figure
\ref{fig:mass_flav} for an explicit example where relations between
the neutrino masses forbid certain areas in parameter space.

\subsection{\label{sec:exotic}Exotic modifications of the three neutrino picture}

There are a number of very exotic modifications to the standard
picture discussed so far, which we will shortly discuss in this
section.

The most obvious modification is that neutrinos are 
{\bf Dirac particles}, in which case there is no \onbb~and reading 
this review was all in vain. Though in fact Dirac neutrinos are the
most straightforward SM extension that can explain massive neutrinos,
the associated Yukawa couplings to generate masses less than eV are at
least 6 orders of magnitude smaller than the electron Yukawa
coupling. For muon and tau neutrinos the coupling is even smaller than
the one of the associated charged lepton. Since the masses in quark
doublets are very close to each other, one considers the extreme
hierarchy in the lepton masses as unnatural and fine-tuned. This is
one of the reasons why the Weinberg operator (\ref{eq:Leff}) and its
explicit realizations in terms of seesaw mechanisms, which suppress 
neutrino mass naturally, are considered as realistic origin of
neutrino mass. 

Dirac neutrinos can be written as two maximally mixed Majorana
neutrinos with common mass $m_i$ and opposite CP parity. The effective mass is then 
\be
\sum\limits_i \sqrt{\frac 12} \, 
|U_{ei}|^2 \, \left( m_i + m_i \, e^{i \pi} \right) = 0 \, . 
\ee
A small splitting of the degeneracy of a state can be described with the mass
matrix
\be
 \hspace{-1cm} m_i 
\left( \baz 
\epsilon & 1 \\
1 & 0 
\ea \right) \rightarrow 
\tilde U = \sqrt{\frac12 } \left( 
\baz 
1 + \frac \epsilon 4 & -1 + \frac \epsilon 4 \\
1 - \frac \epsilon 4 & 1 + \frac \epsilon 4
\ea
\right) \mbox{ and } 
m_i^{\pm} = m_i \left(\pm 1 + \frac \epsilon 2 \right) , 
\ee
with the indicated new eigenstates and mixing (sub-)matrix. 
These {\bf Pseudo-Dirac neutrinos} lead to a very small contribution to the effective
mass of about $\epsilon \, m_i  = \frac 12 \, \delta m^2 /m_i$, with $\delta m^2 =
(m_i^+)^2 - (m_i^-)^2$. See \cite{deGouvea:2009fp} for current 
constraints on $\delta m^2$ from oscillation experiments. If all three states
are Pseudo-Dirac the effective mass is basically zero
\cite{Maalampi:2009wm}. 

If one or two are Pseudo-Dirac interesting predictions for the
effective mass arise. This can happen in ``bimodal'' or {\bf ``schizophrenic''}
scenarios \cite{Allahverdi:2010us}, in which at leading order one or two mass
states are Dirac particles while the other one is Majorana. 
For instance, if $\nu_2$ is a Dirac particle then the
effective mass in the inverted hierarchy is $\meff \simeq \sqrt{\dma}
\, c_{12}^2 \, c_{13}^2$, roughly a factor of two larger than the
minimal  value in the standard case  \cite{Allahverdi:2010us}, see (\ref{eq:meeIHleft}). A
generalization to all possibilities can be found in \cite{Barry:2010en}.

Another exotic property is {\bf CPT violation}. Interesting consequences
for \obb~have been considered in \cite{Barenboim:2002hx}, where a simple
one family example was discussed. Strictly speaking, the neutrinos
cannot fulfill the Majorana condition, but \onbb~still can take
place. 

The situation is slightly different for {\bf tachyonic neutrinos},
whose equation of motion does not allow to define a charge
conjugation, and \obb~has been argued to be absent
\cite{Chodos:1984cy,Jentschura:2012rd}.

\section{\label{sec:sterile}Light sterile neutrinos}

The majority of neutrino experiments is consistent with the 3-neutrino
picture. The noteworthy exception is a number of measurements and
observations in particle physics, astrophysics and cosmology that can
be explained with the presence of light (eV or sub-eV) sterile
neutrinos. See \cite{Abazajian:2012ys} for a review on the current
situation.  Typical values are 
$\Delta m^2_{\rm st} \simeq 1$ eV$^2$, and mixing of order $U_{e4}
\simeq 0.15$ \cite{Kopp:2011qd}. Scenarios with two or more sterile neutrinos
are mildly disfavored by cosmology, since they contribute with two eV-scale
masses to the sum of masses and with additional degrees of freedom
\cite{Hamann:2011ge}.  On the other hand, scenarios with two sterile
neutrinos provide better fits to the anomalous experimental
results. Moreover, while oscillation analysis require typically
mass-squared differences close to eV, cosmology would work better with
less than eV masses. A number of dedicated oscillation experiments
will put the sterile neutrino hypothesis to the test
\cite{Abazajian:2012ys}, and definite answers might be present within
this decade.

 What is the effect for \onbb? This has been discussed several times
 in the past \cite{Bilenky:2001xq,Pakvasa:2002hn,Giunti:2011cp}\footnote{Obviously, such sterile neutrinos can also be tested also in the other
neutrino mass approaches, i.e.~KATRIN and cosmology.}. 
For one sterile neutrino, the effective mass is now a sum of four
terms, the additional term
quantifies the sterile contribution: 
\be 
\meff = | \underbrace{|U_{e1}|^2  m_1 + |U_{e2}|^2  m_2 \, e^{2 i
\alpha} + |U_{e3}^2|  m_3 \, e^{2 i \beta} 
}_{\meff^{\rm act}} + 
\underbrace{|U_{e4}|^2  m_4 \, e^{2 i \Phi_1}}_{\meff^{\rm st}} |  \, .
\ee 
Here $|\meff^{\rm act}|$ is the 3-neutrino contribution discussed so far, and
$\meff^{\rm st}$ the sterile one. 

Note that in case sterile neutrinos are
present, the symmetrical parametrization (\ref{writeout}) has
several advantages \cite{Rodejohann:2011vc}. For instance, if there is
one sterile neutrino there are six independent rotations and six CP
phases (3 Dirac and 3 Majorana). If each
rotation contains a phase, the very often awkward attribution of CP 
phases to the PMNS matrix is not necessary, and taken care of
automatically. 
  
\begin{figure}

\begin{center}
\includegraphics[height=6cm,width=10cm]{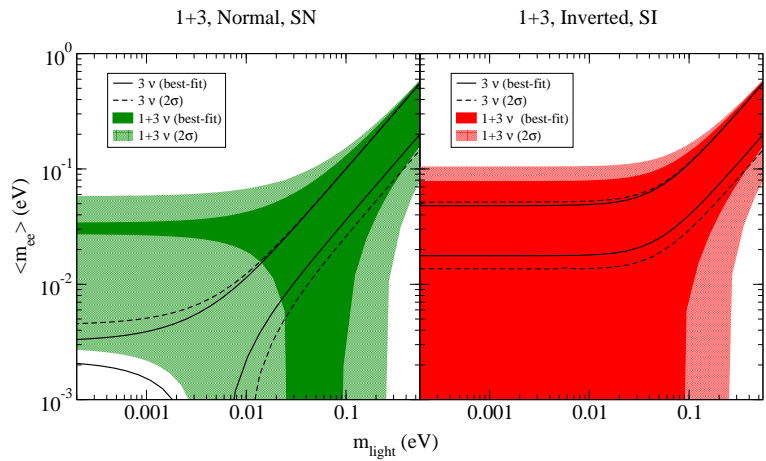}
\end{center}

\vspace*{8pt}
 \caption{Effective mass against the smallest mass in case one sterile
neutrino, heavier than the 3 active ones, is present. The black solid and dashed lines correspond to the
standard 3 neutrino best-fit and $2\sigma$ cases. See \cite{Barry:2011wb}.}
\label{fig:meff_st}
\end{figure}

Suppose the easiest (and least problematic in what regards cosmology) case that there
is only one sterile neutrino, and that it is heavier than the 3 active
ones. The complete list of possibilities including the case of two
sterile neutrinos can be found in
Refs.~\cite{Goswami:2005ng,Goswami:2007kv,Barry:2011wb}. These other 
cases are usually corresponding to quasi-degenerate mass schemes.  
Recall from above that $|\meff_{\rm NH}^{\rm act}|$ can vanish and that 
$|\meff_{\rm IH}^{\rm act}|$ cannot vanish, having a typical value of
0.02 eV. With the typical sterile neutrino parameters given above 
one has 
\be 
|\meff^{\rm st}| \simeq \sqrt{\Delta m^2_{\rm st}} \, |U_{e4}|^2 \simeq 0.02~{\rm eV} \left\{ 
\ba 
\gg |\meff_{\rm NH}^{\rm act} | \, ,\\
\simeq |\meff_{\rm IH}^{\rm act}| \, .
\ea 
\right. 
\ee 
Therefore, if the active neutrinos are normally ordered, the effective
mass {\it cannot} vanish anymore, whereas it {\it can} vanish when they are
inversely ordered \cite{Barry:2011wb}. Hence, the usual standard phenomenology has been
completely turned around! Given that the addition of light sterile
neutrinos is presumably the simplest modification of the standard
picture, this example shows that when discussing the physics potential of
\obb~one should carefully list one's assumptions. 
Figure \ref{fig:meff_st} is a more detailed analysis of the
situation \cite{Barry:2011wb}, in which the smallest mass is plotted against the effective
mass, where the sterile neutrino parameters are from the global fit
results of \cite{Kopp:2011qd}.

\section{\label{sec:concl}Summary}
In this contribution we have focussed on the presumably best motivated
neutrino physics aspect of neutrinoless double beta decay, but there are many other 
scenarios which can lead to \obb, including left-right symmetry,
supersymmetry, extra dimensions, etc. A similar discussion to the one
presented here can be performed
for all of them, involving tests in collider physics, lepton flavor
violation and so on. If a signal in neutrinoless double beta decay is
established in one or more experiment, an exciting physics program
will start, aiming to pin down the underlying mechanism.

It is indeed an exciting time for neutrinoless double beta decay and neutrino
physics. A number of experiments is testing new life-time regimes of
neutrinoless double beta decay, impressively encompassing enormous experimental
difficulties. Theoretical and experimental progress of the nuclear
physics part is also rapidly increasing. 
Combined with the progress in neutrino oscillation
physics, and its further improvement in the years to come, we are 
about to enter new and unexplored regimes. The future will show which
part of the large physics potential can be realized, or if even new
possibilities open up.

\section*{Acknowledgements}
I thank J.~Barry for help in producing figures. 
This work was supported by the ERC under the Starting Grant 
MANITOP. 

\section*{References}

\bibliography{WR}
\bibliographystyle{jphysg}

\end{document}